\documentclass[%
 reprint,
 aps,
 amsmath,amssymb,
 prl]{revtex4-2}
\usepackage{CJK}

\usepackage{mathtools,extarrows,mathrsfs}
\usepackage{float}
\usepackage{subfigure,tikz,tikzscale}
\usetikzlibrary{matrix,arrows.meta,fit,positioning,decorations.markings,shapes.geometric}

\usepackage{titlesec}
\titleformat{\section}{\centering\normalsize\normalfont\bf}{\thesection}{0em}{}
\usepackage{hyperref}
\usepackage{comment}
\usepackage{tabularray}
\usepackage[T1]{fontenc}
\definecolor{darkgreen}{rgb}{0.0, 0.4, 0.0}
\UseTblrLibrary{diagbox}
\usepackage[dvipsnames]{xcolor}

\begin{document}
\begin{CJK*}{UTF8}{}
\CJKfamily{gbsn}
\title{Hidden simplicity in AdS spinning Mellin amplitudes via scaffolding}
\author{Song He (何颂)$^{1,2,3,4}$}
\email{songhe@itp.ac.cn}
\author{Xiang Li (李想)$^{2,3}$}
\email{lixiang@itp.ac.cn}
\author{Yuyu Mo (莫裕宇)$^{2}$}
\email{moyuyu@itp.ac.cn}
\author{Dongyu Yang (杨东昱)$^{1,2,3}$}
\email{yangdongyu24@mails.ucas.ac.cn}
\affiliation{
$^{1}$School of Fundamental Physics and Mathematical Sciences, Hangzhou Institute for Advanced Study \& ICTP-AP, Hangzhou, Zhejiang 310024, China\\
$^{2}$ Institute of Theoretical Physics, Chinese Academy of Sciences, Beijing 100190, China \\
$^{3}$School of Physical Sciences, University of Chinese Academy of Sciences, No.19A Yuquan Road, Beijing 100049, China\\
$^{4}$Peng Huanwu Center for Fundamental Theory, Hefei, Anhui 230026, P. R. China
}
\date{\today}

\begin{abstract}
We uncover surprising hidden simplicity in Mellin amplitudes for tree-level AdS holographic correlators for spinning operators, such as AdS ``gluons'' and ``gravitons'' (spin $1$ and $2$). We define Mellin amplitudes with $n$ spinning operators via the so-called ``scaffolding'' of $2n$-scalar ones with specific projection operators for each spin state, which are rational functions of Mellin variables of $2n$ scalars generalizing flat-space scaffolding amplitudes. We classify possible three-point structures with spin $1$ and $2$ which take the same form as massive three-point amplitudes in flat space, and match with special solutions such as those extracted from $6$-scalar ones in AdS$_5\times S^3$ or AdS$_5\times S^5$. Focusing on AdS$_5$ gluons, we directly bootstrap spinning amplitudes in scaffolding form up to $n=6$ gluons (which amounts to $2n=12$ scalars) using factorizations, multi-linearity and flat-space limit. The results take a remarkably simple form in analogy with flat-space amplitudes, which can be constructed from familiar $3$- and $4$-vertices as well as propagators of massive spin-$1$ particles. Surprisingly, we find that vertices with any descendant levels are proportional to the primary ones with nice combinatorial coefficients, which makes manifest the correct flat-space limit in the simplest possible way. 
\end{abstract}

\maketitle
\end{CJK*}

\section{Introduction}
Recent years have witnessed remarkable progress in computing and revealing new structures of holographic correlators, or ``scattering amplitudes'' in AdS space ({\it c.f.} tree~\cite{Rastelli:2016nze, Rastelli:2017udc,Rastelli:2017ymc,Zhou:2017zaw,Goncalves:2019znr,Alday:2020lbp,Alday:2020dtb,Zhou:2021gnu,Goncalves:2023oyx} and loop~\cite{Alday:2017xua,Aprile:2017bgs,Aprile:2017qoy,Aprile:2019rep,Alday:2019nin,Huang:2021xws,Drummond:2022dxw} level). A natural language for holographic correlators is the Mellin representation~\cite{Mack:2009mi,Penedones:2010ue,Fitzpatrick:2011ia}, where amplitudes factorize at physical poles with the residues encoding the operators exchanged in OPE~\cite{Goncalves:2014rfa}, while the flat-space limit is captured by poles at infinity. Following pioneering works~\cite{Zhou:2018ofp,Alday:2021odx,Alday:2022lkk,Bissi:2022wuh,Alday:2023kfm} in AdS super-Yang-Mills (sYM) theories (see~\cite{Alday:2021ajh,Huang:2023oxf,Huang:2023ppy,Huang:2024rxr} for loop level), a new method has been proposed in~\cite{Cao:2023cwa, Cao:2024bky} for constructing higher-point supergluon amplitudes purely from lower-point ones. Despite the above progress, most results and understandings of Mellin amplitudes have been restricted to cases with at most one spinning operator; beyond that, even the precise definition of Mellin amplitudes with arbitrary number of spinning operators/legs has only been carefully studied very recently in~\cite{Huang:2025ieo}, where one introduces additional Mellin variables $\eta$ to represent the ``polarization'' degrees of freedom.

On the other hand, a new way for representing spinning particles, such as gluons and gravitons, as pairs of ``scaffolding'' scalars for flat-space amplitudes has proved to be very fruitful~\cite{Zeros, Gluons, Arkani-Hamed:2023jry}. It allows one to extract all-loop $n$-gluon amplitudes in Yang-Mills theory from $2n$-scalar amplitudes obtained from a simple kinematical shift of the so-called stringy Tr$\,\phi^3$ amplitudes expressed using (combinatorial) ``curve integrals'' on surfaces~\cite{Arkani-Hamed:2023lbd, Arkani-Hamed:2023mvg, Arkani-Hamed:2017mur, Arkani-Hamed:2019plo, Arkani-Hamed:2019vag, Arkani-Hamed:2019mrd}. Already at tree level, the $n$-gluon scaffolding amplitude becomes a canonical rational function of kinematic variables of $2n$ scalars, where important notions based on gluon polarizations, such as gauge invariance, multi-linearity and factorizations, all have natural combinatorial origins from curves on this simplest surface, the $2n$-gon~\cite{Gluons}. 

\begin{figure}[htbp]
    \includegraphics[scale=0.8]{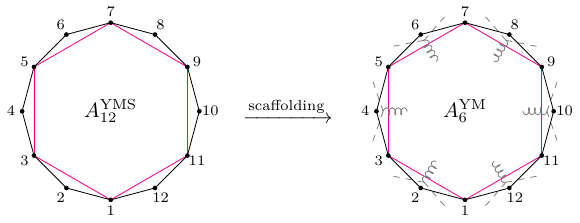}
    \caption{Scaffolding from $2n=12$ scalars to $n=6$ gluons.}
    \label{fig:scaffolding12to6}
\end{figure}

In this Letter, we combine these ideas and propose a new way for representing AdS correlators with arbitrary number of spinning operators in terms of ``scaffolding scalars'': the Mellin amplitudes with $n$ spinning operators can be obtained by acting the $2n$-scalar one with appropriate projection operators. As a warm up exercise, we classify three-point correlators with spin $1$ and $2$, which is equivalent to massive amplitudes in flat space, and we match our results with those from scaffolding $6$-scalar amplitudes of lowest Kaluza--Klein (KK) modes in $\mathcal{N}=1$ sYM on AdS$_5\times S^3$ and  IIB supergravity on AdS$_5\times S^5$ (see Appendix~\hyperref[appendixA]{A} for a review). Focusing on tree-level amplitudes of the Yang-Mills-scalar (YMS) sector in AdS$_5$ background, we then directly bootstrap $n$-gluon amplitudes using factorizations, multi-linearity and flat-space limit (which are nicely the scaffolding Yang-Mills amplitudes~\cite{Arkani-Hamed:2023jry}). As we will see, it is straightforward to construct these amplitudes explicitly through $n=6$, which amounts to the most interesting parts of ``supergluon'' amplitudes up to $2n=12$~\cite{Cao:2023cwa, Cao:2024bky}! Moreover, we reveal hidden simplicity of these new results in scaffolding variables; quite nicely we find that the $n$-gluon amplitudes can be built from Feynman rules with $3$- and $4$-vertices (including all descendant levels), as well as Proca propagators, which are formally identical to massive spin-$1$ particles in flat space. In other words, through $n=6$ we discover a set of ``emergent'' Feynman rules where $3$- and $4$- vertices with any descendant levels are proportional to flat-space ones (in scaffolding variables). Moreover, these descendant coefficients enjoy a nice combinatorial interpretation, and for any Feynman diagram, they conspire to give the same overall coefficient (following from a combinatorial identity), which makes flat-space limit completely manifest!

\section{Spinning correlators from scaffolded and projected Mellin amplitudes}
In this section, we show how to extract $n$-point AdS spinning correlators in Mellin representation from $2n$-scalar ones, in a way very similar to the flat-space case proposed in~\cite{Arkani-Hamed:2023jry}. Recall that it is natural to represent $n$ gluons (same works for gravitons) using $2n$ pairs of scalars {\it e.g.} in $2n$-scalar amplitudes of Yang-Mills-scalar theory, by taking the scaffolding residue ${X}_{1,3}={X}_{3,5}=\cdots={X}_{1,2n-1}=0$, and identifying the gluon kinematic data (momenta $k_i$ and polarizations $e_i$) with linear combinations of scalar momenta $p_{2i-1}, p_{2i}$ via
\begin{equation}\label{ptoEK}
    e_i^\mu \equiv(1-\alpha)p^\mu_{2i}-\alpha p^\mu_{2i-1},\qquad k^\mu_i\equiv p^\mu_{2i}+p^\mu_{2i-1}\,.
\end{equation}
In this way, we fuse $2n$ pairs of colored scalars into $n$ on-shell gluons, as shown in Fig. \ref{fig:scaffolding12to6}.

Our goal is to extend this formalism to $2n$-point Mellin amplitudes of scalar correlators \cite{Mack:2009mi}
\begin{equation}\label{def:Mellin}
    G_{12 \cdots 2n}=\int[{\rm d}\delta]\;\mathcal M_{2n}(\delta_{ij})\prod_{i<j}\frac{\Gamma(\delta_{ij})}{(-2P_i\cdot P_j)^{\delta_{ij}}}.
\end{equation}
While the scaffolded spinning amplitude is independent of details of the $2n$ scalars, we employ the scalar operators with conformal dimension $\Delta = 2$ for simplicity. Here we use the embedding formalism following~\cite{Goncalves:2014rfa}, where $P_i\cdot P_j=-\frac12(x_i-x_j)^2$. The Mellin variables are constrained as $\delta_{ij}=p_i\cdot p_j$ for auxiliary momenta satisfying $\sum_i p_i=0$ and $\delta_{ii}=p_i^2=-\tau_i=-2$, with conformal twist $\tau_i:=\Delta_i-J_i$ ($J_i$ is the spin of an operator). Inspired by discussions in flat space~\cite{Arkani-Hamed:2017mur}, we also introduce the $\tfrac{2n(2n-3)}{2}$ planar variables
\begin{equation}
{\bf X}_{i,j}\equiv-1-\frac{1}{2}\sum_{i\leq k, l<j} \delta_{kl}=-1-\frac{1}{2}\left(\sum_{i\leq k<j}p_k\right)^2,
\end{equation}
where we have ${\bf X}_{i,j}\equiv{\bf X}_{j,i}$, ${\bf X}_{i,i{+}1}=0$ and ${\bf X}_{i,i}= -1$, with indices understood modulo $2n$. The inverse transform, motivated by the associahedron in~\cite{Arkani-Hamed:2017mur, Arkani-Hamed:2019vag}, reads:
\begin{equation}
    \delta_{ij}={\bf X}_{i,j}+{\bf X}_{i+1,j+1}-{\bf X}_{i,j+1}-{\bf X}_{i+1,j}\;.
\end{equation}

\begin{figure}[htbp]
    \centering
    \includegraphics[scale=0.8]{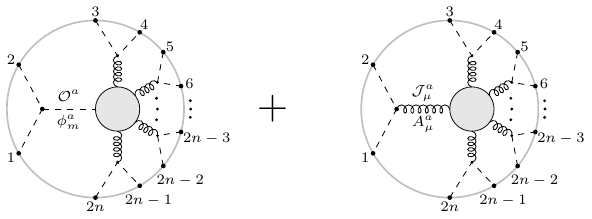}
    \caption{Mixture of spin-0 (left) and spin-1 (right) contribution in scaffolding residue on $\mathbf{X}_{1,3}$.}
    \label{fig:mixture}
\end{figure}
With the planar variables in hand, we now proceed to extract the scaffolded $n$-spinning amplitudes from the $2n$-scalar Mellin amplitudes by taking the residue at ${\bf X}_{2i-1, 2i+1} = 0$ for $i=1,\dots,n$, leaving $2n(n-2)$ independent variables. 
However, unlike flat-space YMS where scaffolding cleanly isolates the $n$-gluon amplitude, the AdS scaffolding residue generally mixes exchange channels from cubic couplings of the scaffolded scalar pairs to bulk fields of various spins: scalars, vectors, or higher spin tensors (see {\it e.g.} Fig.~\ref{fig:mixture} for a mixture of spin-0,1 exchanges). To address this issue, we employ spin projectors to disentangle these contributions by selecting terms with specific powers of ${\bf e}_i$. Motivated by the flat-space multi-linearity condition~\cite{Arkani-Hamed:2023jry}, we define $\mathcal{P}^{({\bf e})}_i$ as the AdS analogue of ${\bf e}_i \cdot \partial_{{\bf e}_i}$ acting on the $i$-th external leg: 
\begin{equation}
    \mathcal{P}^{({\bf e})}_{i} = \sum_{j \neq 2i, 2i\pm 1} \left( \mathbf{X}_{2i,j} - \frac{\mathbf{X}_{2i-1, j} + \mathbf{X}_{2i+1, j} + 1}{2} \right) \frac{\partial}{\partial \mathbf{X}_{2i,j}}\,,
    \label{def:projector}
\end{equation}
where we have adopted the choice $\alpha=1/2$ here and throughout the rest of this work. The operator $\mathcal{P}^{({\bf e})}_i$ serves as the fundamental building block for the general spin-$\ell$ projector $\mathcal{P}^{(\ell)}$, whose precise form is determined by the maximum spin involved in the amplitudes. For the simplest case of spins up to 1, $\mathcal{P}^{({\bf e})}_i$ acts as ${\bf e}_i \cdot \partial_{{\bf e}_i}$ to extract the term linear in ``polarization'' ${\bf e}_i$ (see Appendix~\hyperref[appendixB]{B}), implying $\mathcal{P}^{(1)}_i = \mathcal{P}^{({\bf e})}_i$ and $\mathcal{P}^{(0)}_i=1-\mathcal{P}^{(\mathbf{e})}_i$. For the case of spins up to 2, the projectors can be constructed as polynomials in $\mathcal{P}^{({\bf e})}$ in a similar manner, as shown in Table~\ref{tab:spin2_projectors}.
\begin{table}[H]
\centering
\begin{tblr}{rowspec={|[1pt]Q[c,m]|Q[c,m]Q[c,m]Q[c,m]|[1pt]},colspec={Q[l,m]Q[c,m]}}
    component&projector\\
    graviton&$\mathcal{P}^{(2)}=\frac{1}{2}\left((\mathcal{P}^{(\mathbf{e})})^2-\mathcal{P}^{(\mathbf{e})}\right)$\\
    photon&$\mathcal{P}^{(1)}=2\mathcal{P}^{(\mathbf{e})}-(\mathcal{P}^{(\mathbf{e})})^2$\\
    scalar&$\mathcal{P}^{(0)}=1-\frac{3}{2}\mathcal{P}^{(\mathbf{e})}+\frac{1}{2}(\mathcal{P}^{(\mathbf{e})})^2$
\end{tblr}
\caption{Projectors for extracting specific spin components from the scaffolded scalar amplitude involving spin-0,1,2.}
\label{tab:spin2_projectors}
\end{table}

The spinning amplitude $ \mathcal{A}_n^{(s_1,\dots ,s_n)} $ for arbitrary spins follows from the formula:
\begin{equation}
      \mathcal{A}_n^{(s_1, \dots, s_n)}\equiv\prod_{i=1}^n\mathcal{P}_i^{(s_i)}\underset{\mathbf{X}_{2i-1,2i+1}=0}{\operatorname{Res}}\mathcal{M}_{2n}\,.
      \label{eq:general_spin_amp}
\end{equation}
Note that in general, one may also define spinning amplitudes analogous to~\eqref{eq:general_spin_amp}, but with residues taken from descendant poles. Here we restrict to the case of primary spinning correlators based on available data~\cite{Cao:2023cwa,Goncalves:2025jcg}. 

\section{Three-point amplitudes}

As a warm up, we determine possible structures of three-point spinning amplitudes with $s=1,2$, which turns out to be in one-to-one correspondence with those for massive three-point amplitudes in flat space. We impose two conditions, the \textbf{multi-linearity}, {\it i.e.} 
\begin{equation}
	\prod_{i=1}^n \mathcal{P}_i^{\left(s_i\right)}\mathcal{A}^{\left(s_1, \dots,  s_n\right)}_n=\mathcal{A}^{\left(s_1, \dots, s_n\right)}_n ,\qquad s_i=1 \text{ or } 2,
\end{equation}
as well as {\bf flat-space limit}, {\it i.e.} with rescaling $\mathbf{X}\to \beta \mathbf{X}$, AdS spinning amplitudes must reproduce flat-space amplitudes in the limit $\beta\to\infty$. For simplicity, we focus on the leading power-counting case, namely the flat-space limit   corresponds to Yang-Mills for spins $(1,1,1)$ and General Relativity for spins $(2,2,2)$ (we also elaborate on amplitudes with mixed spins in the Appendix~\hyperref[appendixA]{A}). 

For $s_i=1$, we construct an ansatz of a degree-$2$ polynomial of ${\bf X}$ variables; imposing both conditions fixes the answer (up to an overall constant):
\begin{equation}
\begin{aligned}
	\mathcal{A}^{(1,1,1)}_3 = A_3 + \sum_{i=1}^{3}\left[\mathbf{X}_{2i,2i+2}-\frac{1}{2}\mathbf{X}_{i,i+3}\right] - \frac{3}{4},\\
\end{aligned}
\label{resultccc}
\end{equation}
where $A_3 = 2\sum_{i=1}^{3} \mathbf{X}_{i,i+3} \left(\mathbf{X}_{i+1,i+4}-\mathbf{X}_{2-2i,4-2i}\right)$ is the leading quadratic term matching the flat space scaffolded Yang-Mills three-point amplitude~\cite{Arkani-Hamed:2023jry}, and the linear and constant terms are completely fixed by multi-linearity.

For $(2,2,2)$, starting from a degree-$4$ polynomial, these two constrains leave one coefficient unfixed (in addition to the overall constant): 
\begin{equation}
\begin{aligned}
	\mathcal{A}^{(2,2,2)}_3 =& \left(\mathcal{A}^{(1,1,1)}_3\right)^2
    +\lambda~{\bf y}_1 {\bf y}_2 {\bf y}_3 ,
\end{aligned}
\label{resultcccc}
\end{equation}
where the first term survives in flat space limit and reproduces three-graviton amplitude (double copy of three-gluon one, $(A_3)^2$), ${\bf y}_i= \mathbf{X}_{i,i+3}+\mathbf{X}_{i+1,i+4}-2 \mathbf{X}_{6-2i,4-2i}+\tfrac{1}{2}$ and $\lambda$ a theory-dependent, unfixed coefficient. It is satisfying to see that three-point spinning correlators with $s=1,2$ are almost fixed by these conditions, and become beautiful generalizations of flat-space scaffolding amplitudes. Note that the number of solutions depends on the power-counting of the ansatz and the flat-space limit: if we use a degree-$3$ ansatz for the $(1,1,1)$ case, thereby admitting a Tr$\,F^3$ term in the flat-space limit, we would find two solutions, matching the results of~\cite{Huang:2025ieo}.

In fact, with a simple change of variables, we can make these results formally identical to flat-space amplitudes with massive spinning particles! Analogous to the flat-space scaffolding gluon kinematic~\eqref{ptoEK}, we introduce formal momenta ${\bf k}_i$ and polarizations ${\bf e}_i$ with $\alpha=1/2$, together with their ``contractions'' defined as:
\begin{align}
&\begin{aligned}
    {\bf ee}_{ij}&=\frac{1}{4}(\delta_{2i,2j}+\delta_{2i-1,2j-1}-\delta_{2i,2j-1}-\delta_{2i-1,2j}),\\
    {\bf ek}_{ij}&=\frac{1}{2}(\delta_{2i,2j}-\delta_{2i-1,2j-1}+\delta_{2i,2j-1}-\delta_{2i-1,2j}),\\
    {\bf kk}_{ij}&=\delta_{2i,2j}+\delta_{2i-1,2j-1}+\delta_{2i,2j-1}+\delta_{2i-1,2j},
\end{aligned}
\label{eq:delta2EK}
\end{align}
where we still have ``transversality'' ${\bf ek}_{ii}=0$ and massive, ``on-shell'' conditions, ${\bf ee}_{ii}=-\frac{3}{2}$ and ${\bf kk}_{ii}=-2$. We also have the analogue of ``momentum conservation'' $\sum_{j} \mathbf{ek}_{ij} =0$ and $\sum_{j} \mathbf{kk}_{ij}=0$. It is easy to count that the number of variables is still $2n(n{-}2)$. \eqref{eq:delta2EK} goes back to the familiar transformation between $e_i \cdot e_j$, $e_i \cdot k_j$, $k_i \cdot k_j$ and $\delta_{i,j}$ in the flat-space limit where those boundary values of ${\bf ee}_{ii}$, ${\bf kk}_{ii}$ all go to zero as $\delta_{ii}$ goes to zero.

Surprisingly, the scaffolding amplitudes become homogeneous functions of these formal polarizations ${\bf e}_i$. For $n=3$, \eqref{resultccc} becomes formally identical to the three-gluon amplitude, or more precisely, the amplitude for three massive spin-$1$ particles: 
\begin{equation}
    \mathcal{A}_3^{(1,1,1)} =
    2 \left({\bf ee}_{12} {\bf ek}_{31}+{\bf ee}_{23} {\bf ek}_{12} +{\bf ee}_{31} {\bf ek}_{23}   \right),
    \label{spin1EK}
\end{equation}
Similarly, \eqref{resultcccc} matches exactly the two possible structures for the three-point spin-$2$ amplitude in dRGT massive gravity~\cite{deRham:2010kj,Momeni:2020vvr}
\begin{equation}
\begin{aligned}
    \mathcal{A}_3^{(2,2,2)}=& 4 \left ({\bf ee}_{12} {\bf ek}_{31}+{\bf ee}_{23} {\bf ek}_{12} +{\bf ee}_{31} {\bf ek}_{23} \right)^2\\
    &-8\;   \lambda \; {\bf ee}_{12} {\bf ee}_{23} {\bf ee}_{31}.
\end{aligned}
\label{spin2EK}
\end{equation}
Again, the first term is the double copy of \eqref{spin1EK}, while the second is also well known in massive gravity~\cite{Momeni:2020vvr}. Very nicely, the former agrees with three-gluon amplitudes in AdS$_5\times S^3$ (extracted from $6$-scalar results in~\cite{Alday:2023kfm, Cao:2023cwa}), and by extracting residues from the tour de force calculation of $6$-scalar amplitude in AdS$_5\times S^5$ supergravity~\cite{Goncalves:2025jcg} we find that it corresponds to \eqref{spin2EK} with $\lambda=1$. 

We emphasize that \eqref{spin1EK} and \eqref{spin2EK} are general results for $s_i=1$ and $s_i=2$ cases (which also generalize to other dimensions). As ``massive amplitudes'', they do not respect gauge invariance under ``${\bf e}_i \to {\bf e}_i + \beta {\bf k}_i$'', though such invariance is recovered in the flat-space limit~\cite{Arkani-Hamed:2023jry}.

\section{Bootstrap \texorpdfstring{$n$}{n}-gluon amplitudes and Feynman rules}
Next we focus on the Yang-Mills-scalar sector of $\mathcal{N}=1$ sYM in AdS$_5\times S^3$, with the lowest KK modes as scalars: we show how to bootstrap $n$-gluon amplitudes and reveal hidden simplicity in our explicit results up to $n=6$. 
\paragraph{Bootstrap} We denote $\mathcal{A}_n^{\mathrm{YMS}}=\mathcal{A}_n^{\left(1,\dots,1\right)}$, and in addition to the two conditions, {\bf multi-linearity} and {\bf flat-space limit}, we also impose \textbf{factorizations}; the exchange of spin-1 operators dictates that ${\cal A}_n^{\rm YMS}$ factories on residues of primary and descendant poles~\cite{Costa:2014kfa}:
\begin{equation}\label{eq:YMS-fac}
	\underset{{\mathbf{X}}_{1,2j+1}=m}{\mathrm{Res}}	\mathcal{A}_n^{\mathrm{YMS}}\sim f(\mathbf{X},m)
	\partial_ {\mathbf{X}_{i, I}}{\mathcal{A}_{L}^{\mathrm{YMS}\;(m)}}
	\partial_ {\mathbf{X}_{j, I^\prime}}\mathcal{A}_{R}^{\mathrm{YMS}\;(m)},
\end{equation}
where $m=0,1,2,\dots$ denotes the descendant level whose truncation was studied in {\it e.g.} \cite{Cao:2024bky}; $I$, $I^\prime$ denote the exchange indices and $i$ denote the indices carried by the amplitude on the left and $j$ on the right. The precise formula is discussed in Appendix~\hyperref[appendixB]{B}.

We use four-point amplitude as an illustrative example which has the following structure
\begin{equation}
	\mathcal{A}_{4}^{\rm YMS}=\frac{A_{15}}{\mathbf{X}_{1,5}}+\frac{B_{15}}{\mathbf{X}_{1,5}-1}+\frac{A_{37}}{\mathbf{X}_{3,7}}+\frac{B_{37}}{\mathbf{X}_{3,7}-1}+R_4.
\end{equation}
The $A$'s and $B$'s are completely determined by factorization where we glue two $\mathcal{A}_3^{(1,1,1)}$ together.
To determine $R_4$, the flat-space limit fixes the highest power quadratic terms in $\mathbf{X}$ while multi-linearity fixes the lower-power terms.
The result is given compactly in \eqref{4pt_res}. The same procedure extends straightforwardly to all $n$, which can be done without any difficulty at least up to $n=6$. Nevertheless, they correspond to the most complicated and interesting part of ``supergluon'' amplitudes with $2n\leqslant 12$ scalars, which had been worked out completely only up to $2n=8$ in the literature \cite{Cao:2023cwa}.

\paragraph{Emergent Feynman rules}
Recall that we introduced a change of variables in \eqref{eq:delta2EK} which renders the three-point expressions particularly simple. These variables not only trivialize the projector (${\cal P}_i^{(\bf e)} \sim {\bf e}_i \cdot \partial/\partial{\bf e}_i$), but they also make ``spinning'' factorizations, flat-space limit {\it etc.} manifest. We show that, much like their flat-space counterparts, AdS gluon amplitudes exhibit hidden simplicity and novel structures. 

We find that a remarkable set of ``Feynman rules'' emerge from our explicit results: not only have we found no ``contact'' vertices beyond four points (all those contact terms in ${\bf X}$ variables are gone), but the ``effective'' $3$- and $4$-vertices all take identical form as the flat-space ones! We denote these ``contracted'' vertices as:
\begin{eqnarray}\label{eq:V3pt4pt}
&V(123)
\equiv{\bf ee}_{12}({\bf ek}_{31}-{\bf ek} _{32})+\text{cyclic},    \nonumber\\
&V(1234)
\equiv -2 {\bf ee}_{13} {\bf ee}_{24}+{\bf ee}_{12} {\bf ee}_{34}+{\bf ee}_{14} {\bf ee}_{23},
\end{eqnarray}
and a highly-nontrivial observation through $n=6$ (which we conjecture to hold for any $n$) is that $3$- and $4$-vertices with any {\it descendant levels}, $\{m_1,\dots, m_k\}$ for $k=3,4$, take the same form and the only dependence on $\{m_i\}$ is the overall coefficient defined by $V_{m_1, m_2, m_3}(123)\equiv\alpha_{m_1,m_2,m_3}V(123)$ and $V_{m_1,\dots, m_4}(1234)\equiv\alpha_{m_1,\dots, m_4}V(1234)$. We find, for primary ones, $\alpha_{000}=2$, $\alpha_{0000}=6$, and for descendant cases needed up to $n=6$, $\alpha_{001}=\alpha_{012}=-1, \alpha_{011}=2, \alpha_{111}=-4$, and $\alpha_{0001}=-6, \alpha_{0011}=12, \alpha_{0002}=2$; we conjecture that the coefficients $\alpha_{m_1,m_2,\dots, m_k}$ for $k=3,4$ read:
\begin{equation}
\sum_{n_1=0}^{m_1}\cdots\sum_{n_k=0}^{m_k}\Gamma\left(\sum_{a=1}^k n_a +k\right) \prod_{a=1}^k \frac{(-m_a)_{n_a}}{n_a! (n_a{+}1)!} \,,
\end{equation}
with the Pochhammer symbol $(a)_b \equiv \Gamma(a+b)/\Gamma(a)$. 
Remarkably, these coefficients exactly match Witten-diagram computations in~\cite{Li:2023azu}, and we present a nice recursion relation for $\alpha_{m_1,m_2,\dots, m_k}$ in Appendix~\hyperref[appendixC]{C}. 

\begin{figure}
    \centering
    \includegraphics[scale=1.0]{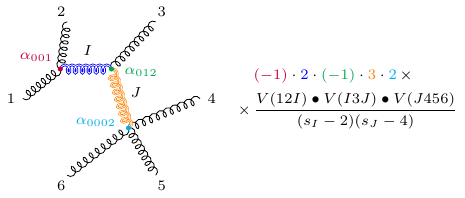}
    \caption{An example of 6-pt Feynman diagram ($3*3*4$ channel) where the two propagators have descendant levels $m_I=1$ and $m_J=2$, respectively.}
    \label{fig:a_6pt_example}
\end{figure}

When gluing two vertices ${V}_L(\cdots,I)$ and ${V}_R(I,\cdots)$ on an internal leg $I$, it is understood as if ${\bf ee}, {\bf ek}$ and ${\bf kk}$ come from contractions of ``vectors'' ${\bf e}^\mu$ and ${\bf k}^\mu$, multiplied by a ``Proca'' propagator (at level $m_I=0,1,\dots$), {\it i.e.}
\begin{equation}
    {V}_L\bullet{V}_R\equiv\frac{\partial{V}_L}{\partial\mathbf e_I^\mu}\left(\eta^{\mu\nu}+\frac{{\bf k}_I^\mu {\bf k}_I^\nu}{2+2m_{I}}\right)\frac{\partial{V}_R}{\partial\mathbf e_I^\nu},
\end{equation}
where ${\bf k}_I^\mu\equiv -\sum_{i\in L} {\bf k}_i^\mu=\sum_{i\in R} {\bf k}_i^\mu$ denotes the ``momentum'' of $I$, and the derivative w.r.t. ``polarization'' ${\bf e}_I^\mu$ is via those w.r.t. ${\bf ee}_{I, i}, {\bf ek}_{I,i}$.  For example, when two $3$-vertices are glued, $V(12 I) \bullet V(I34)$, the contraction gives
\begin{equation}
\begin{aligned}
	4\big[&\mathbf{ee}_{12} \left(\mathbf{ek}_{31} \mathbf{ek}_{42}-\mathbf{ek}_{32} \mathbf{ek}_{41}\right)+(12 \leftrightarrow 34) \\\nonumber
	+ &\mathbf{ek}_{12} \left(\mathbf{ee}_{24} \mathbf{ek}_{34}-\mathbf{ee}_{23} \mathbf{ek}_{43}\right)+(13 \leftrightarrow 24)	\big]\\\nonumber
	&\qquad\qquad-2 {\bf ee}_{12} {\bf ee}_{34} (2{\bf kk}_{23}+{\bf kk}_{12}-2).
\label{v3v3}    
\end{aligned}    
\end{equation}
At least through $n=6$, the $n$-gluon amplitudes exactly follow from these Feynman rules including all descendants, which take the form (see Fig.~\ref{fig:a_6pt_example}):
\begin{equation}
\begin{aligned}\label{eq18}
\mathcal A^{\rm YMS}_n =\sum_{g} \sum_{\{m\}} \frac{N_{g,\{m\}}} {\prod_{I\in E(g)} D_I^{(m_I)}}\hspace{2cm}\\
=\sum_{g}  \sum_{\{m\}}(V \bullet V \bullet \cdots \bullet V) \frac{\prod_{v\in V(g)} \alpha_{\{m_v\}}} {\prod_{I\in E(g)} D_I^{(m_I)}},
\end{aligned}
\end{equation}
where we sum over all planar tree graphs with $3$- and $4$-vertices~\footnote{The total number of such graphs is nicely given in \url{https://oeis.org/A001002}.}, $g$, as well as possible descendant levels for all propagators $I$, $\{m_I\}$; the poles are normalized according to the convention~\cite{Li:2023azu}: $D_I^{(m_I)}\equiv\frac{s_I-  2m_I}{m_I{+}1}$ with $s_I\equiv 2 \mathbf{X}_{2i{-}1,2j{-}1}$ for $I=\{i,i{+}1,\dots, j{-}1\}$~\footnote{In general, $s_{I}\equiv-2-\sum_{a,b\in I}{\bf kk}_{ab}$ for an arbitrary subset $I$ of indices.}, and the numerator $N_{g,\{m\}}$ obtained from gluing $V$'s together; the dependence of $N_{g, \{m\}}$ on descendant levels is only through Proca propagators in the gluing and the product of $\alpha$'s for vertices $v$~\footnote{For each vertex we have three possible levels, $m_v^{(1,2,3)}$, which are given by descendant levels on the corresponding propagators.}! For example, the four-point amplitude is a sum of two cubic graphs and a quartic one:
\begin{equation}
\begin{aligned}
\mathcal A^{\rm YMS}_4= & V(12 I) \bullet V(I 34) \left(\frac{2^2}{s_{12}} + \frac{(-1)^2\times 2}{ s_{12}-2}\right)\\ &+(i \to i{+}1)  + 6\times V(1234)\,,
\end{aligned}
\label{4pt_res}
\end{equation}
and the five-point amplitude consists of $5+5$ graphs~\footnote{For $n=4,5$, every propagator connects to a cubic vertex with two external legs, so the contraction $V \bullet V$ can be equivalently performed with $\eta^{\mu \nu}$. However, this is no longer true for some graphs starting from $n=6$.}:
\begin{widetext}
\begin{equation}
\begin{aligned}
\mathcal A^{\rm YMS}_5= & V(12I)\bullet V(I3J) \bullet V(J45) \left(\frac{2^3}{s_{12} s_{45}} + \frac{(-1)^2\times 2^2}{(s_{12}{-}2) s_{45}} + \frac{(-1)^2\times2^2}{ s_{12} (s_{45}{-}2)} +\frac{(-1)^2\times2^3}{(s_{12}{-}2) (s_{45}{-}2)} \right)+{\rm cyclic}
\\
&+  V(12 I) \bullet V(I345) \left(\frac{6\times2}{s_{12}} + \frac{(-6)\times(-1)\times2}{s_{12}{-}2} \right)+{\rm cyclic}.
\end{aligned}
\label{5pt_res}
\end{equation}
\end{widetext}
For $n=6$, there are $38$ graphs from seven cyclic seeds as we enumerate in the Appendix~\hyperref[appendixC]{C} (see Fig.~\ref{fig:a_6pt_example} for an example). Very nicely, this make the flat-space limit completely manifest, where \eqref{eq:V3pt4pt} reduce to flat-space vertices and the sum of scalar graphs (with descendant levels) in~\eqref{eq18} reduces to the flat-space graph times a coefficient made of products of $\alpha$'s!
For any graph of $n=4,5,6$, the products of $\alpha$'s indeed conspire to give $3!$, $4!$ and $5!$, respectively; as we show in Appendix~\hyperref[appendixC]{C}, the magic behind this is the beautiful recursion relation, $\alpha_{m_1\dots m_n}=\sum_{m_I=0}^\infty\alpha_{m_1\dots m_km_I}(m_I+1)\alpha_{m_Im_{k+1}\dots m_{n}}$ (and the fact that it is crossing symmetric), which guarantees that any Feynman diagram picks the {\it same} overall coefficient in the flat-space limit, which is $\alpha_{0,\dots, 0}=(n{-}1)!$. 

\section{Outlook}
In this Letter, we have initiated a study on Mellin amplitudes for AdS spinning correlators by scaffolding the scalar ones similar to that for flat-space amplitudes. We set up the formalism and study general structures of three-point spinning amplitudes, which we then match with AdS amplitudes of ``gluons'' and ``gravitons'' from scaffolding. Focusing on Yang-Mills-scalar theory, we then bootstrap $n$-gluon amplitudes based on flat-space limit, multi-linearity and factorizations. Remarkably we find an emergent set of Feynman rules of massive spin-$1$ particles in scaffolding variables, where vertices with different descendant levels all take identical form up to nice coefficients with combinatorial interpretations. 

Our preliminary investigations open up numerous directions to explore. It would be highly desirable to derive Feynman rules for all descendant levels directly, and find more structures underlying $n$-gluon amplitudes such as color-kinematics duality~\cite{Bern:2008qj,Bern:2010ue} and possible ``universal expansions''~\cite{Dong:2021qai}. Practically it would also be advantageous to have direct recursion relations generalizing the flat-space ones~\cite{Berends:1987me,Britto:2005fq}. Could there be a ``curve-integral'' formulation for these scaffolding gluon amplitudes~\cite{Arkani-Hamed:2023lbd,Arkani-Hamed:2023mvg}, and even for string amplitudes in AdS~\cite{Aprile:2020luw,Abl:2020dbx,Aprile:2020mus,Aprile:2022tzr,Alday:2022uxp,Alday:2022xwz,Alday:2023jdk,Alday:2023mvu,Wang:2025pjo}? 
Moving beyond gluons, it would be interesting to {\it e.g.} bootstrap $n\geqslant 4$ amplitudes in supergravity (which amounts to the most interesting part with $2n\geqslant 8$ scalars) and unveil their underlying structures such as double copy~\cite{Zhou:2021gnu}. It would also be highly desirable to study general Feynman rules including arbitrary KK modes, for AdS spinning amplitudes in and beyond super-Yang-Mills and supergravity theories, where we expect hidden symmetries~\cite{Loebbert:2018xce,Caron-Huot:2018kta,Aprile:2020mus,Aprile:2022tzr,Abl:2020dbx,Caron-Huot:2021usw,Caron-Huot:2023wdh,Aprile:2020luw,Aprile:2022tzr,Huang:2024dxr,Wang:2025pjo,Fernandes:2025eqe}.

\section*{Acknowledgments}
It is our pleasure to thank Yongxiang Su for collaborations in early stages of the work, and Simon Caron-Huot, Zhongjie Huang, Jiajie Mei, Yichao Tang and Xinan Zhou for inspiring discussions or comments on the manuscript. This work has been supported by the National Natural Science Foundation of China under Grant No. 12225510,
12447101, 12247103, and by the New Cornerstone Science
Foundation.

\bibliographystyle{apsrev4-1}
\bibliography{Refs}

\begin{thebibliography}{71}%
\makeatletter
\providecommand \@ifxundefined [1]{%
 \@ifx{#1\undefined}
}%
\providecommand \@ifnum [1]{%
 \ifnum #1\expandafter \@firstoftwo
 \else \expandafter \@secondoftwo
 \fi
}%
\providecommand \@ifx [1]{%
 \ifx #1\expandafter \@firstoftwo
 \else \expandafter \@secondoftwo
 \fi
}%
\providecommand \natexlab [1]{#1}%
\providecommand \enquote  [1]{``#1''}%
\providecommand \bibnamefont  [1]{#1}%
\providecommand \bibfnamefont [1]{#1}%
\providecommand \citenamefont [1]{#1}%
\providecommand \href@noop [0]{\@secondoftwo}%
\providecommand \href [0]{\begingroup \@sanitize@url \@href}%
\providecommand \@href[1]{\@@startlink{#1}\@@href}%
\providecommand \@@href[1]{\endgroup#1\@@endlink}%
\providecommand \@sanitize@url [0]{\catcode `\\12\catcode `\$12\catcode `\&12\catcode `\#12\catcode `\^12\catcode `\_12\catcode `\%12\relax}%
\providecommand \@@startlink[1]{}%
\providecommand \@@endlink[0]{}%
\providecommand \url  [0]{\begingroup\@sanitize@url \@url }%
\providecommand \@url [1]{\endgroup\@href {#1}{\urlprefix }}%
\providecommand \urlprefix  [0]{URL }%
\providecommand \Eprint [0]{\href }%
\providecommand \doibase [0]{http://dx.doi.org/}%
\providecommand \selectlanguage [0]{\@gobble}%
\providecommand \bibinfo  [0]{\@secondoftwo}%
\providecommand \bibfield  [0]{\@secondoftwo}%
\providecommand \translation [1]{[#1]}%
\providecommand \BibitemOpen [0]{}%
\providecommand \bibitemStop [0]{}%
\providecommand \bibitemNoStop [0]{.\EOS\space}%
\providecommand \EOS [0]{\spacefactor3000\relax}%
\providecommand \BibitemShut  [1]{\csname bibitem#1\endcsname}%
\let\auto@bib@innerbib\@empty
\bibitem [{\citenamefont {Rastelli}\ and\ \citenamefont {Zhou}(2017)}]{Rastelli:2016nze}%
  \BibitemOpen
  \bibfield  {author} {\bibinfo {author} {\bibfnamefont {L.}~\bibnamefont {Rastelli}}\ and\ \bibinfo {author} {\bibfnamefont {X.}~\bibnamefont {Zhou}},\ }\href {\doibase 10.1103/PhysRevLett.118.091602} {\bibfield  {journal} {\bibinfo  {journal} {Phys. Rev. Lett.}\ }\textbf {\bibinfo {volume} {118}},\ \bibinfo {pages} {091602} (\bibinfo {year} {2017})},\ \Eprint {http://arxiv.org/abs/1608.06624} {arXiv:1608.06624 [hep-th]} \BibitemShut {NoStop}%
\bibitem [{\citenamefont {Rastelli}\ and\ \citenamefont {Zhou}(2018{\natexlab{a}})}]{Rastelli:2017udc}%
  \BibitemOpen
  \bibfield  {author} {\bibinfo {author} {\bibfnamefont {L.}~\bibnamefont {Rastelli}}\ and\ \bibinfo {author} {\bibfnamefont {X.}~\bibnamefont {Zhou}},\ }\href {\doibase 10.1007/JHEP04(2018)014} {\bibfield  {journal} {\bibinfo  {journal} {JHEP}\ }\textbf {\bibinfo {volume} {04}},\ \bibinfo {pages} {014} (\bibinfo {year} {2018}{\natexlab{a}})},\ \Eprint {http://arxiv.org/abs/1710.05923} {arXiv:1710.05923 [hep-th]} \BibitemShut {NoStop}%
\bibitem [{\citenamefont {Rastelli}\ and\ \citenamefont {Zhou}(2018{\natexlab{b}})}]{Rastelli:2017ymc}%
  \BibitemOpen
  \bibfield  {author} {\bibinfo {author} {\bibfnamefont {L.}~\bibnamefont {Rastelli}}\ and\ \bibinfo {author} {\bibfnamefont {X.}~\bibnamefont {Zhou}},\ }\href {\doibase 10.1007/JHEP06(2018)087} {\bibfield  {journal} {\bibinfo  {journal} {JHEP}\ }\textbf {\bibinfo {volume} {06}},\ \bibinfo {pages} {087} (\bibinfo {year} {2018}{\natexlab{b}})},\ \Eprint {http://arxiv.org/abs/1712.02788} {arXiv:1712.02788 [hep-th]} \BibitemShut {NoStop}%
\bibitem [{\citenamefont {Zhou}(2018{\natexlab{a}})}]{Zhou:2017zaw}%
  \BibitemOpen
  \bibfield  {author} {\bibinfo {author} {\bibfnamefont {X.}~\bibnamefont {Zhou}},\ }\href {\doibase 10.1007/JHEP08(2018)187} {\bibfield  {journal} {\bibinfo  {journal} {JHEP}\ }\textbf {\bibinfo {volume} {08}},\ \bibinfo {pages} {187} (\bibinfo {year} {2018}{\natexlab{a}})},\ \Eprint {http://arxiv.org/abs/1712.02800} {arXiv:1712.02800 [hep-th]} \BibitemShut {NoStop}%
\bibitem [{\citenamefont {Gon\c{c}alves}\ \emph {et~al.}(2019)\citenamefont {Gon\c{c}alves}, \citenamefont {Pereira},\ and\ \citenamefont {Zhou}}]{Goncalves:2019znr}%
  \BibitemOpen
  \bibfield  {author} {\bibinfo {author} {\bibfnamefont {V.}~\bibnamefont {Gon\c{c}alves}}, \bibinfo {author} {\bibfnamefont {R.}~\bibnamefont {Pereira}}, \ and\ \bibinfo {author} {\bibfnamefont {X.}~\bibnamefont {Zhou}},\ }\href {\doibase 10.1007/JHEP10(2019)247} {\bibfield  {journal} {\bibinfo  {journal} {JHEP}\ }\textbf {\bibinfo {volume} {10}},\ \bibinfo {pages} {247} (\bibinfo {year} {2019})},\ \Eprint {http://arxiv.org/abs/1906.05305} {arXiv:1906.05305 [hep-th]} \BibitemShut {NoStop}%
\bibitem [{\citenamefont {Alday}\ and\ \citenamefont {Zhou}(2020{\natexlab{a}})}]{Alday:2020lbp}%
  \BibitemOpen
  \bibfield  {author} {\bibinfo {author} {\bibfnamefont {L.~F.}\ \bibnamefont {Alday}}\ and\ \bibinfo {author} {\bibfnamefont {X.}~\bibnamefont {Zhou}},\ }\href {\doibase 10.1103/PhysRevLett.125.131604} {\bibfield  {journal} {\bibinfo  {journal} {Phys. Rev. Lett.}\ }\textbf {\bibinfo {volume} {125}},\ \bibinfo {pages} {131604} (\bibinfo {year} {2020}{\natexlab{a}})},\ \Eprint {http://arxiv.org/abs/2006.06653} {arXiv:2006.06653 [hep-th]} \BibitemShut {NoStop}%
\bibitem [{\citenamefont {Alday}\ and\ \citenamefont {Zhou}(2021)}]{Alday:2020dtb}%
  \BibitemOpen
  \bibfield  {author} {\bibinfo {author} {\bibfnamefont {L.~F.}\ \bibnamefont {Alday}}\ and\ \bibinfo {author} {\bibfnamefont {X.}~\bibnamefont {Zhou}},\ }\href {\doibase 10.1103/PhysRevX.11.011056} {\bibfield  {journal} {\bibinfo  {journal} {Phys. Rev. X}\ }\textbf {\bibinfo {volume} {11}},\ \bibinfo {pages} {011056} (\bibinfo {year} {2021})},\ \Eprint {http://arxiv.org/abs/2006.12505} {arXiv:2006.12505 [hep-th]} \BibitemShut {NoStop}%
\bibitem [{\citenamefont {Zhou}(2021)}]{Zhou:2021gnu}%
  \BibitemOpen
  \bibfield  {author} {\bibinfo {author} {\bibfnamefont {X.}~\bibnamefont {Zhou}},\ }\href {\doibase 10.1103/PhysRevLett.127.141601} {\bibfield  {journal} {\bibinfo  {journal} {Phys. Rev. Lett.}\ }\textbf {\bibinfo {volume} {127}},\ \bibinfo {pages} {141601} (\bibinfo {year} {2021})},\ \Eprint {http://arxiv.org/abs/2106.07651} {arXiv:2106.07651 [hep-th]} \BibitemShut {NoStop}%
\bibitem [{\citenamefont {Gon\c{c}alves}\ \emph {et~al.}(2023)\citenamefont {Gon\c{c}alves}, \citenamefont {Meneghelli}, \citenamefont {Pereira}, \citenamefont {Vilas~Boas},\ and\ \citenamefont {Zhou}}]{Goncalves:2023oyx}%
  \BibitemOpen
  \bibfield  {author} {\bibinfo {author} {\bibfnamefont {V.}~\bibnamefont {Gon\c{c}alves}}, \bibinfo {author} {\bibfnamefont {C.}~\bibnamefont {Meneghelli}}, \bibinfo {author} {\bibfnamefont {R.}~\bibnamefont {Pereira}}, \bibinfo {author} {\bibfnamefont {J.}~\bibnamefont {Vilas~Boas}}, \ and\ \bibinfo {author} {\bibfnamefont {X.}~\bibnamefont {Zhou}},\ }\href {\doibase 10.1007/JHEP08(2023)067} {\bibfield  {journal} {\bibinfo  {journal} {JHEP}\ }\textbf {\bibinfo {volume} {08}},\ \bibinfo {pages} {067} (\bibinfo {year} {2023})},\ \Eprint {http://arxiv.org/abs/2302.01896} {arXiv:2302.01896 [hep-th]} \BibitemShut {NoStop}%
\bibitem [{\citenamefont {Alday}\ and\ \citenamefont {Bissi}(2017)}]{Alday:2017xua}%
  \BibitemOpen
  \bibfield  {author} {\bibinfo {author} {\bibfnamefont {L.~F.}\ \bibnamefont {Alday}}\ and\ \bibinfo {author} {\bibfnamefont {A.}~\bibnamefont {Bissi}},\ }\href {\doibase 10.1103/PhysRevLett.119.171601} {\bibfield  {journal} {\bibinfo  {journal} {Phys. Rev. Lett.}\ }\textbf {\bibinfo {volume} {119}},\ \bibinfo {pages} {171601} (\bibinfo {year} {2017})},\ \Eprint {http://arxiv.org/abs/1706.02388} {arXiv:1706.02388 [hep-th]} \BibitemShut {NoStop}%
\bibitem [{\citenamefont {Aprile}\ \emph {et~al.}(2018{\natexlab{a}})\citenamefont {Aprile}, \citenamefont {Drummond}, \citenamefont {Heslop},\ and\ \citenamefont {Paul}}]{Aprile:2017bgs}%
  \BibitemOpen
  \bibfield  {author} {\bibinfo {author} {\bibfnamefont {F.}~\bibnamefont {Aprile}}, \bibinfo {author} {\bibfnamefont {J.~M.}\ \bibnamefont {Drummond}}, \bibinfo {author} {\bibfnamefont {P.}~\bibnamefont {Heslop}}, \ and\ \bibinfo {author} {\bibfnamefont {H.}~\bibnamefont {Paul}},\ }\href {\doibase 10.1007/JHEP01(2018)035} {\bibfield  {journal} {\bibinfo  {journal} {JHEP}\ }\textbf {\bibinfo {volume} {01}},\ \bibinfo {pages} {035} (\bibinfo {year} {2018}{\natexlab{a}})},\ \Eprint {http://arxiv.org/abs/1706.02822} {arXiv:1706.02822 [hep-th]} \BibitemShut {NoStop}%
\bibitem [{\citenamefont {Aprile}\ \emph {et~al.}(2018{\natexlab{b}})\citenamefont {Aprile}, \citenamefont {Drummond}, \citenamefont {Heslop},\ and\ \citenamefont {Paul}}]{Aprile:2017qoy}%
  \BibitemOpen
  \bibfield  {author} {\bibinfo {author} {\bibfnamefont {F.}~\bibnamefont {Aprile}}, \bibinfo {author} {\bibfnamefont {J.~M.}\ \bibnamefont {Drummond}}, \bibinfo {author} {\bibfnamefont {P.}~\bibnamefont {Heslop}}, \ and\ \bibinfo {author} {\bibfnamefont {H.}~\bibnamefont {Paul}},\ }\href {\doibase 10.1007/JHEP05(2018)056} {\bibfield  {journal} {\bibinfo  {journal} {JHEP}\ }\textbf {\bibinfo {volume} {05}},\ \bibinfo {pages} {056} (\bibinfo {year} {2018}{\natexlab{b}})},\ \Eprint {http://arxiv.org/abs/1711.03903} {arXiv:1711.03903 [hep-th]} \BibitemShut {NoStop}%
\bibitem [{\citenamefont {Aprile}\ \emph {et~al.}(2020)\citenamefont {Aprile}, \citenamefont {Drummond}, \citenamefont {Heslop},\ and\ \citenamefont {Paul}}]{Aprile:2019rep}%
  \BibitemOpen
  \bibfield  {author} {\bibinfo {author} {\bibfnamefont {F.}~\bibnamefont {Aprile}}, \bibinfo {author} {\bibfnamefont {J.}~\bibnamefont {Drummond}}, \bibinfo {author} {\bibfnamefont {P.}~\bibnamefont {Heslop}}, \ and\ \bibinfo {author} {\bibfnamefont {H.}~\bibnamefont {Paul}},\ }\href {\doibase 10.1007/JHEP03(2020)190} {\bibfield  {journal} {\bibinfo  {journal} {JHEP}\ }\textbf {\bibinfo {volume} {03}},\ \bibinfo {pages} {190} (\bibinfo {year} {2020})},\ \Eprint {http://arxiv.org/abs/1912.01047} {arXiv:1912.01047 [hep-th]} \BibitemShut {NoStop}%
\bibitem [{\citenamefont {Alday}\ and\ \citenamefont {Zhou}(2020{\natexlab{b}})}]{Alday:2019nin}%
  \BibitemOpen
  \bibfield  {author} {\bibinfo {author} {\bibfnamefont {L.~F.}\ \bibnamefont {Alday}}\ and\ \bibinfo {author} {\bibfnamefont {X.}~\bibnamefont {Zhou}},\ }\href {\doibase 10.1007/JHEP09(2020)008} {\bibfield  {journal} {\bibinfo  {journal} {JHEP}\ }\textbf {\bibinfo {volume} {09}},\ \bibinfo {pages} {008} (\bibinfo {year} {2020}{\natexlab{b}})},\ \Eprint {http://arxiv.org/abs/1912.02663} {arXiv:1912.02663 [hep-th]} \BibitemShut {NoStop}%
\bibitem [{\citenamefont {Huang}\ and\ \citenamefont {Yuan}(2023)}]{Huang:2021xws}%
  \BibitemOpen
  \bibfield  {author} {\bibinfo {author} {\bibfnamefont {Z.}~\bibnamefont {Huang}}\ and\ \bibinfo {author} {\bibfnamefont {E.~Y.}\ \bibnamefont {Yuan}},\ }\href {\doibase 10.1007/JHEP04(2023)064} {\bibfield  {journal} {\bibinfo  {journal} {JHEP}\ }\textbf {\bibinfo {volume} {04}},\ \bibinfo {pages} {064} (\bibinfo {year} {2023})},\ \Eprint {http://arxiv.org/abs/2112.15174} {arXiv:2112.15174 [hep-th]} \BibitemShut {NoStop}%
\bibitem [{\citenamefont {Drummond}\ and\ \citenamefont {Paul}(2022)}]{Drummond:2022dxw}%
  \BibitemOpen
  \bibfield  {author} {\bibinfo {author} {\bibfnamefont {J.~M.}\ \bibnamefont {Drummond}}\ and\ \bibinfo {author} {\bibfnamefont {H.}~\bibnamefont {Paul}},\ }\href {\doibase 10.1007/JHEP08(2022)275} {\bibfield  {journal} {\bibinfo  {journal} {JHEP}\ }\textbf {\bibinfo {volume} {08}},\ \bibinfo {pages} {275} (\bibinfo {year} {2022})},\ \Eprint {http://arxiv.org/abs/2204.01829} {arXiv:2204.01829 [hep-th]} \BibitemShut {NoStop}%
\bibitem [{\citenamefont {Mack}(2009)}]{Mack:2009mi}%
  \BibitemOpen
  \bibfield  {author} {\bibinfo {author} {\bibfnamefont {G.}~\bibnamefont {Mack}},\ }\href@noop {} {\  (\bibinfo {year} {2009})},\ \Eprint {http://arxiv.org/abs/0907.2407} {arXiv:0907.2407 [hep-th]} \BibitemShut {NoStop}%
\bibitem [{\citenamefont {Penedones}(2011)}]{Penedones:2010ue}%
  \BibitemOpen
  \bibfield  {author} {\bibinfo {author} {\bibfnamefont {J.}~\bibnamefont {Penedones}},\ }\href {\doibase 10.1007/JHEP03(2011)025} {\bibfield  {journal} {\bibinfo  {journal} {JHEP}\ }\textbf {\bibinfo {volume} {03}},\ \bibinfo {pages} {025} (\bibinfo {year} {2011})},\ \Eprint {http://arxiv.org/abs/1011.1485} {arXiv:1011.1485 [hep-th]} \BibitemShut {NoStop}%
\bibitem [{\citenamefont {Fitzpatrick}\ \emph {et~al.}(2011)\citenamefont {Fitzpatrick}, \citenamefont {Kaplan}, \citenamefont {Penedones}, \citenamefont {Raju},\ and\ \citenamefont {van Rees}}]{Fitzpatrick:2011ia}%
  \BibitemOpen
  \bibfield  {author} {\bibinfo {author} {\bibfnamefont {A.~L.}\ \bibnamefont {Fitzpatrick}}, \bibinfo {author} {\bibfnamefont {J.}~\bibnamefont {Kaplan}}, \bibinfo {author} {\bibfnamefont {J.}~\bibnamefont {Penedones}}, \bibinfo {author} {\bibfnamefont {S.}~\bibnamefont {Raju}}, \ and\ \bibinfo {author} {\bibfnamefont {B.~C.}\ \bibnamefont {van Rees}},\ }\href {\doibase 10.1007/JHEP11(2011)095} {\bibfield  {journal} {\bibinfo  {journal} {JHEP}\ }\textbf {\bibinfo {volume} {11}},\ \bibinfo {pages} {095} (\bibinfo {year} {2011})},\ \Eprint {http://arxiv.org/abs/1107.1499} {arXiv:1107.1499 [hep-th]} \BibitemShut {NoStop}%
\bibitem [{\citenamefont {Gon\c{c}alves}\ \emph {et~al.}(2015)\citenamefont {Gon\c{c}alves}, \citenamefont {Penedones},\ and\ \citenamefont {Trevisani}}]{Goncalves:2014rfa}%
  \BibitemOpen
  \bibfield  {author} {\bibinfo {author} {\bibfnamefont {V.}~\bibnamefont {Gon\c{c}alves}}, \bibinfo {author} {\bibfnamefont {J.}~\bibnamefont {Penedones}}, \ and\ \bibinfo {author} {\bibfnamefont {E.}~\bibnamefont {Trevisani}},\ }\href {\doibase 10.1007/JHEP10(2015)040} {\bibfield  {journal} {\bibinfo  {journal} {JHEP}\ }\textbf {\bibinfo {volume} {10}},\ \bibinfo {pages} {040} (\bibinfo {year} {2015})},\ \Eprint {http://arxiv.org/abs/1410.4185} {arXiv:1410.4185 [hep-th]} \BibitemShut {NoStop}%
\bibitem [{\citenamefont {Zhou}(2018{\natexlab{b}})}]{Zhou:2018ofp}%
  \BibitemOpen
  \bibfield  {author} {\bibinfo {author} {\bibfnamefont {X.}~\bibnamefont {Zhou}},\ }\href {\doibase 10.1007/JHEP07(2018)147} {\bibfield  {journal} {\bibinfo  {journal} {JHEP}\ }\textbf {\bibinfo {volume} {07}},\ \bibinfo {pages} {147} (\bibinfo {year} {2018}{\natexlab{b}})},\ \Eprint {http://arxiv.org/abs/1804.02397} {arXiv:1804.02397 [hep-th]} \BibitemShut {NoStop}%
\bibitem [{\citenamefont {Alday}\ \emph {et~al.}(2021)\citenamefont {Alday}, \citenamefont {Behan}, \citenamefont {Ferrero},\ and\ \citenamefont {Zhou}}]{Alday:2021odx}%
  \BibitemOpen
  \bibfield  {author} {\bibinfo {author} {\bibfnamefont {L.~F.}\ \bibnamefont {Alday}}, \bibinfo {author} {\bibfnamefont {C.}~\bibnamefont {Behan}}, \bibinfo {author} {\bibfnamefont {P.}~\bibnamefont {Ferrero}}, \ and\ \bibinfo {author} {\bibfnamefont {X.}~\bibnamefont {Zhou}},\ }\href {\doibase 10.1007/JHEP06(2021)020} {\bibfield  {journal} {\bibinfo  {journal} {JHEP}\ }\textbf {\bibinfo {volume} {06}},\ \bibinfo {pages} {020} (\bibinfo {year} {2021})},\ \Eprint {http://arxiv.org/abs/2103.15830} {arXiv:2103.15830 [hep-th]} \BibitemShut {NoStop}%
\bibitem [{\citenamefont {Alday}\ \emph {et~al.}(2022{\natexlab{a}})\citenamefont {Alday}, \citenamefont {Gon\c{c}alves},\ and\ \citenamefont {Zhou}}]{Alday:2022lkk}%
  \BibitemOpen
  \bibfield  {author} {\bibinfo {author} {\bibfnamefont {L.~F.}\ \bibnamefont {Alday}}, \bibinfo {author} {\bibfnamefont {V.}~\bibnamefont {Gon\c{c}alves}}, \ and\ \bibinfo {author} {\bibfnamefont {X.}~\bibnamefont {Zhou}},\ }\href {\doibase 10.1103/PhysRevLett.128.161601} {\bibfield  {journal} {\bibinfo  {journal} {Phys. Rev. Lett.}\ }\textbf {\bibinfo {volume} {128}},\ \bibinfo {pages} {161601} (\bibinfo {year} {2022}{\natexlab{a}})},\ \Eprint {http://arxiv.org/abs/2201.04422} {arXiv:2201.04422 [hep-th]} \BibitemShut {NoStop}%
\bibitem [{\citenamefont {Bissi}\ \emph {et~al.}(2023)\citenamefont {Bissi}, \citenamefont {Fardelli}, \citenamefont {Manenti},\ and\ \citenamefont {Zhou}}]{Bissi:2022wuh}%
  \BibitemOpen
  \bibfield  {author} {\bibinfo {author} {\bibfnamefont {A.}~\bibnamefont {Bissi}}, \bibinfo {author} {\bibfnamefont {G.}~\bibnamefont {Fardelli}}, \bibinfo {author} {\bibfnamefont {A.}~\bibnamefont {Manenti}}, \ and\ \bibinfo {author} {\bibfnamefont {X.}~\bibnamefont {Zhou}},\ }\href {\doibase 10.1007/JHEP01(2023)021} {\bibfield  {journal} {\bibinfo  {journal} {JHEP}\ }\textbf {\bibinfo {volume} {01}},\ \bibinfo {pages} {021} (\bibinfo {year} {2023})},\ \Eprint {http://arxiv.org/abs/2209.01204} {arXiv:2209.01204 [hep-th]} \BibitemShut {NoStop}%
\bibitem [{\citenamefont {Alday}\ \emph {et~al.}(2024)\citenamefont {Alday}, \citenamefont {Gon{\c{c}}alves}, \citenamefont {Nocchi},\ and\ \citenamefont {Zhou}}]{Alday:2023kfm}%
  \BibitemOpen
  \bibfield  {author} {\bibinfo {author} {\bibfnamefont {L.~F.}\ \bibnamefont {Alday}}, \bibinfo {author} {\bibfnamefont {V.}~\bibnamefont {Gon{\c{c}}alves}}, \bibinfo {author} {\bibfnamefont {M.}~\bibnamefont {Nocchi}}, \ and\ \bibinfo {author} {\bibfnamefont {X.}~\bibnamefont {Zhou}},\ }\href {\doibase 10.1103/PhysRevResearch.6.L012041} {\bibfield  {journal} {\bibinfo  {journal} {Phys. Rev. Res.}\ }\textbf {\bibinfo {volume} {6}},\ \bibinfo {pages} {L012041} (\bibinfo {year} {2024})},\ \Eprint {http://arxiv.org/abs/2307.06884} {arXiv:2307.06884 [hep-th]} \BibitemShut {NoStop}%
\bibitem [{\citenamefont {Alday}\ \emph {et~al.}(2022{\natexlab{b}})\citenamefont {Alday}, \citenamefont {Bissi},\ and\ \citenamefont {Zhou}}]{Alday:2021ajh}%
  \BibitemOpen
  \bibfield  {author} {\bibinfo {author} {\bibfnamefont {L.~F.}\ \bibnamefont {Alday}}, \bibinfo {author} {\bibfnamefont {A.}~\bibnamefont {Bissi}}, \ and\ \bibinfo {author} {\bibfnamefont {X.}~\bibnamefont {Zhou}},\ }\href {\doibase 10.1007/JHEP02(2022)105} {\bibfield  {journal} {\bibinfo  {journal} {JHEP}\ }\textbf {\bibinfo {volume} {02}},\ \bibinfo {pages} {105} (\bibinfo {year} {2022}{\natexlab{b}})},\ \Eprint {http://arxiv.org/abs/2110.09861} {arXiv:2110.09861 [hep-th]} \BibitemShut {NoStop}%
\bibitem [{\citenamefont {Huang}\ \emph {et~al.}(2023)\citenamefont {Huang}, \citenamefont {Wang}, \citenamefont {Yuan},\ and\ \citenamefont {Zhou}}]{Huang:2023oxf}%
  \BibitemOpen
  \bibfield  {author} {\bibinfo {author} {\bibfnamefont {Z.}~\bibnamefont {Huang}}, \bibinfo {author} {\bibfnamefont {B.}~\bibnamefont {Wang}}, \bibinfo {author} {\bibfnamefont {E.~Y.}\ \bibnamefont {Yuan}}, \ and\ \bibinfo {author} {\bibfnamefont {X.}~\bibnamefont {Zhou}},\ }\href {\doibase 10.1007/JHEP07(2023)053} {\bibfield  {journal} {\bibinfo  {journal} {JHEP}\ }\textbf {\bibinfo {volume} {07}},\ \bibinfo {pages} {053} (\bibinfo {year} {2023})},\ \Eprint {http://arxiv.org/abs/2301.13240} {arXiv:2301.13240 [hep-th]} \BibitemShut {NoStop}%
\bibitem [{\citenamefont {Huang}\ \emph {et~al.}(2024)\citenamefont {Huang}, \citenamefont {Wang}, \citenamefont {Yuan},\ and\ \citenamefont {Zhou}}]{Huang:2023ppy}%
  \BibitemOpen
  \bibfield  {author} {\bibinfo {author} {\bibfnamefont {Z.}~\bibnamefont {Huang}}, \bibinfo {author} {\bibfnamefont {B.}~\bibnamefont {Wang}}, \bibinfo {author} {\bibfnamefont {E.~Y.}\ \bibnamefont {Yuan}}, \ and\ \bibinfo {author} {\bibfnamefont {X.}~\bibnamefont {Zhou}},\ }\href {\doibase 10.1007/JHEP01(2024)190} {\bibfield  {journal} {\bibinfo  {journal} {JHEP}\ }\textbf {\bibinfo {volume} {01}},\ \bibinfo {pages} {190} (\bibinfo {year} {2024})},\ \Eprint {http://arxiv.org/abs/2309.14413} {arXiv:2309.14413 [hep-th]} \BibitemShut {NoStop}%
\bibitem [{\citenamefont {Huang}\ \emph {et~al.}(2025{\natexlab{a}})\citenamefont {Huang}, \citenamefont {Wang},\ and\ \citenamefont {Yuan}}]{Huang:2024rxr}%
  \BibitemOpen
  \bibfield  {author} {\bibinfo {author} {\bibfnamefont {Z.}~\bibnamefont {Huang}}, \bibinfo {author} {\bibfnamefont {B.}~\bibnamefont {Wang}}, \ and\ \bibinfo {author} {\bibfnamefont {E.~Y.}\ \bibnamefont {Yuan}},\ }\href {\doibase 10.1103/PhysRevLett.134.051601} {\bibfield  {journal} {\bibinfo  {journal} {Phys. Rev. Lett.}\ }\textbf {\bibinfo {volume} {134}},\ \bibinfo {pages} {051601} (\bibinfo {year} {2025}{\natexlab{a}})},\ \Eprint {http://arxiv.org/abs/2407.03408} {arXiv:2407.03408 [hep-th]} \BibitemShut {NoStop}%
\bibitem [{\citenamefont {Cao}\ \emph {et~al.}(2024{\natexlab{a}})\citenamefont {Cao}, \citenamefont {He},\ and\ \citenamefont {Tang}}]{Cao:2023cwa}%
  \BibitemOpen
  \bibfield  {author} {\bibinfo {author} {\bibfnamefont {Q.}~\bibnamefont {Cao}}, \bibinfo {author} {\bibfnamefont {S.}~\bibnamefont {He}}, \ and\ \bibinfo {author} {\bibfnamefont {Y.}~\bibnamefont {Tang}},\ }\href {\doibase 10.1103/PhysRevLett.133.021605} {\bibfield  {journal} {\bibinfo  {journal} {Phys. Rev. Lett.}\ }\textbf {\bibinfo {volume} {133}},\ \bibinfo {pages} {021605} (\bibinfo {year} {2024}{\natexlab{a}})},\ \Eprint {http://arxiv.org/abs/2312.15484} {arXiv:2312.15484 [hep-th]} \BibitemShut {NoStop}%
\bibitem [{\citenamefont {Cao}\ \emph {et~al.}(2024{\natexlab{b}})\citenamefont {Cao}, \citenamefont {He}, \citenamefont {Li},\ and\ \citenamefont {Tang}}]{Cao:2024bky}%
  \BibitemOpen
  \bibfield  {author} {\bibinfo {author} {\bibfnamefont {Q.}~\bibnamefont {Cao}}, \bibinfo {author} {\bibfnamefont {S.}~\bibnamefont {He}}, \bibinfo {author} {\bibfnamefont {X.}~\bibnamefont {Li}}, \ and\ \bibinfo {author} {\bibfnamefont {Y.}~\bibnamefont {Tang}},\ }\href {\doibase 10.1007/JHEP10(2024)040} {\bibfield  {journal} {\bibinfo  {journal} {JHEP}\ }\textbf {\bibinfo {volume} {10}},\ \bibinfo {pages} {040} (\bibinfo {year} {2024}{\natexlab{b}})},\ \Eprint {http://arxiv.org/abs/2406.08538} {arXiv:2406.08538 [hep-th]} \BibitemShut {NoStop}%
\bibitem [{\citenamefont {Huang}\ and\ \citenamefont {Tang}(2025)}]{Huang:2025ieo}%
  \BibitemOpen
  \bibfield  {author} {\bibinfo {author} {\bibfnamefont {Z.}~\bibnamefont {Huang}}\ and\ \bibinfo {author} {\bibfnamefont {Y.}~\bibnamefont {Tang}},\ }\href@noop {} {\  (\bibinfo {year} {2025})},\ \Eprint {http://arxiv.org/abs/2510.07388} {arXiv:2510.07388 [hep-th]} \BibitemShut {NoStop}%
\bibitem [{\citenamefont {Arkani-Hamed}\ \emph {et~al.}(2024)\citenamefont {Arkani-Hamed}, \citenamefont {Cao}, \citenamefont {Dong}, \citenamefont {Figueiredo},\ and\ \citenamefont {He}}]{Zeros}%
  \BibitemOpen
  \bibfield  {author} {\bibinfo {author} {\bibfnamefont {N.}~\bibnamefont {Arkani-Hamed}}, \bibinfo {author} {\bibfnamefont {Q.}~\bibnamefont {Cao}}, \bibinfo {author} {\bibfnamefont {J.}~\bibnamefont {Dong}}, \bibinfo {author} {\bibfnamefont {C.}~\bibnamefont {Figueiredo}}, \ and\ \bibinfo {author} {\bibfnamefont {S.}~\bibnamefont {He}},\ }\href {\doibase 10.1007/JHEP10(2024)231} {\bibfield  {journal} {\bibinfo  {journal} {JHEP}\ }\textbf {\bibinfo {volume} {10}},\ \bibinfo {pages} {231} (\bibinfo {year} {2024})},\ \Eprint {http://arxiv.org/abs/2312.16282} {arXiv:2312.16282 [hep-th]} \BibitemShut {NoStop}%
\bibitem [{\citenamefont {Arkani-Hamed}\ \emph {et~al.}(2025{\natexlab{a}})\citenamefont {Arkani-Hamed}, \citenamefont {Cao}, \citenamefont {Dong}, \citenamefont {Figueiredo},\ and\ \citenamefont {He}}]{Gluons}%
  \BibitemOpen
  \bibfield  {author} {\bibinfo {author} {\bibfnamefont {N.}~\bibnamefont {Arkani-Hamed}}, \bibinfo {author} {\bibfnamefont {Q.}~\bibnamefont {Cao}}, \bibinfo {author} {\bibfnamefont {J.}~\bibnamefont {Dong}}, \bibinfo {author} {\bibfnamefont {C.}~\bibnamefont {Figueiredo}}, \ and\ \bibinfo {author} {\bibfnamefont {S.}~\bibnamefont {He}},\ }\href {\doibase 10.1007/JHEP04(2025)078} {\bibfield  {journal} {\bibinfo  {journal} {JHEP}\ }\textbf {\bibinfo {volume} {04}},\ \bibinfo {pages} {078} (\bibinfo {year} {2025}{\natexlab{a}})},\ \Eprint {http://arxiv.org/abs/2401.00041} {arXiv:2401.00041 [hep-th]} \BibitemShut {NoStop}%
\bibitem [{\citenamefont {Arkani-Hamed}\ \emph {et~al.}(2025{\natexlab{b}})\citenamefont {Arkani-Hamed}, \citenamefont {Cao}, \citenamefont {Dong}, \citenamefont {Figueiredo},\ and\ \citenamefont {He}}]{Arkani-Hamed:2023jry}%
  \BibitemOpen
  \bibfield  {author} {\bibinfo {author} {\bibfnamefont {N.}~\bibnamefont {Arkani-Hamed}}, \bibinfo {author} {\bibfnamefont {Q.}~\bibnamefont {Cao}}, \bibinfo {author} {\bibfnamefont {J.}~\bibnamefont {Dong}}, \bibinfo {author} {\bibfnamefont {C.}~\bibnamefont {Figueiredo}}, \ and\ \bibinfo {author} {\bibfnamefont {S.}~\bibnamefont {He}},\ }\href {\doibase 10.1007/JHEP04(2025)078} {\bibfield  {journal} {\bibinfo  {journal} {JHEP}\ }\textbf {\bibinfo {volume} {04}},\ \bibinfo {pages} {078} (\bibinfo {year} {2025}{\natexlab{b}})},\ \Eprint {http://arxiv.org/abs/2401.00041} {arXiv:2401.00041 [hep-th]} \BibitemShut {NoStop}%
\bibitem [{\citenamefont {Arkani-Hamed}\ \emph {et~al.}(2025{\natexlab{c}})\citenamefont {Arkani-Hamed}, \citenamefont {Frost}, \citenamefont {Salvatori}, \citenamefont {Plamondon},\ and\ \citenamefont {Thomas}}]{Arkani-Hamed:2023lbd}%
  \BibitemOpen
  \bibfield  {author} {\bibinfo {author} {\bibfnamefont {N.}~\bibnamefont {Arkani-Hamed}}, \bibinfo {author} {\bibfnamefont {H.}~\bibnamefont {Frost}}, \bibinfo {author} {\bibfnamefont {G.}~\bibnamefont {Salvatori}}, \bibinfo {author} {\bibfnamefont {P.-G.}\ \bibnamefont {Plamondon}}, \ and\ \bibinfo {author} {\bibfnamefont {H.}~\bibnamefont {Thomas}},\ }\href {\doibase 10.1007/JHEP08(2025)194} {\bibfield  {journal} {\bibinfo  {journal} {JHEP}\ }\textbf {\bibinfo {volume} {08}},\ \bibinfo {pages} {194} (\bibinfo {year} {2025}{\natexlab{c}})},\ \Eprint {http://arxiv.org/abs/2309.15913} {arXiv:2309.15913 [hep-th]} \BibitemShut {NoStop}%
\bibitem [{\citenamefont {Arkani-Hamed}\ \emph {et~al.}(2025{\natexlab{d}})\citenamefont {Arkani-Hamed}, \citenamefont {Frost}, \citenamefont {Salvatori}, \citenamefont {Plamondon},\ and\ \citenamefont {Thomas}}]{Arkani-Hamed:2023mvg}%
  \BibitemOpen
  \bibfield  {author} {\bibinfo {author} {\bibfnamefont {N.}~\bibnamefont {Arkani-Hamed}}, \bibinfo {author} {\bibfnamefont {H.}~\bibnamefont {Frost}}, \bibinfo {author} {\bibfnamefont {G.}~\bibnamefont {Salvatori}}, \bibinfo {author} {\bibfnamefont {P.-G.}\ \bibnamefont {Plamondon}}, \ and\ \bibinfo {author} {\bibfnamefont {H.}~\bibnamefont {Thomas}},\ }\href {\doibase 10.1007/JHEP08(2025)194} {\bibfield  {journal} {\bibinfo  {journal} {JHEP}\ }\textbf {\bibinfo {volume} {08}},\ \bibinfo {pages} {194} (\bibinfo {year} {2025}{\natexlab{d}})},\ \Eprint {http://arxiv.org/abs/2311.09284} {arXiv:2311.09284 [hep-th]} \BibitemShut {NoStop}%
\bibitem [{\citenamefont {Arkani-Hamed}\ \emph {et~al.}(2018)\citenamefont {Arkani-Hamed}, \citenamefont {Bai}, \citenamefont {He},\ and\ \citenamefont {Yan}}]{Arkani-Hamed:2017mur}%
  \BibitemOpen
  \bibfield  {author} {\bibinfo {author} {\bibfnamefont {N.}~\bibnamefont {Arkani-Hamed}}, \bibinfo {author} {\bibfnamefont {Y.}~\bibnamefont {Bai}}, \bibinfo {author} {\bibfnamefont {S.}~\bibnamefont {He}}, \ and\ \bibinfo {author} {\bibfnamefont {G.}~\bibnamefont {Yan}},\ }\href {\doibase 10.1007/JHEP05(2018)096} {\bibfield  {journal} {\bibinfo  {journal} {JHEP}\ }\textbf {\bibinfo {volume} {05}},\ \bibinfo {pages} {096} (\bibinfo {year} {2018})},\ \Eprint {http://arxiv.org/abs/1711.09102} {arXiv:1711.09102 [hep-th]} \BibitemShut {NoStop}%
\bibitem [{\citenamefont {Arkani-Hamed}\ \emph {et~al.}(2023)\citenamefont {Arkani-Hamed}, \citenamefont {He}, \citenamefont {Lam},\ and\ \citenamefont {Thomas}}]{Arkani-Hamed:2019plo}%
  \BibitemOpen
  \bibfield  {author} {\bibinfo {author} {\bibfnamefont {N.}~\bibnamefont {Arkani-Hamed}}, \bibinfo {author} {\bibfnamefont {S.}~\bibnamefont {He}}, \bibinfo {author} {\bibfnamefont {T.}~\bibnamefont {Lam}}, \ and\ \bibinfo {author} {\bibfnamefont {H.}~\bibnamefont {Thomas}},\ }\href {\doibase 10.1103/PhysRevD.107.066015} {\bibfield  {journal} {\bibinfo  {journal} {Phys. Rev. D}\ }\textbf {\bibinfo {volume} {107}},\ \bibinfo {pages} {066015} (\bibinfo {year} {2023})},\ \Eprint {http://arxiv.org/abs/1912.11764} {arXiv:1912.11764 [hep-th]} \BibitemShut {NoStop}%
\bibitem [{\citenamefont {Arkani-Hamed}\ \emph {et~al.}(2022)\citenamefont {Arkani-Hamed}, \citenamefont {He}, \citenamefont {Salvatori},\ and\ \citenamefont {Thomas}}]{Arkani-Hamed:2019vag}%
  \BibitemOpen
  \bibfield  {author} {\bibinfo {author} {\bibfnamefont {N.}~\bibnamefont {Arkani-Hamed}}, \bibinfo {author} {\bibfnamefont {S.}~\bibnamefont {He}}, \bibinfo {author} {\bibfnamefont {G.}~\bibnamefont {Salvatori}}, \ and\ \bibinfo {author} {\bibfnamefont {H.}~\bibnamefont {Thomas}},\ }\href {\doibase 10.1007/JHEP11(2022)049} {\bibfield  {journal} {\bibinfo  {journal} {JHEP}\ }\textbf {\bibinfo {volume} {11}},\ \bibinfo {pages} {049} (\bibinfo {year} {2022})},\ \Eprint {http://arxiv.org/abs/1912.12948} {arXiv:1912.12948 [hep-th]} \BibitemShut {NoStop}%
\bibitem [{\citenamefont {Arkani-Hamed}\ \emph {et~al.}(2021)\citenamefont {Arkani-Hamed}, \citenamefont {He},\ and\ \citenamefont {Lam}}]{Arkani-Hamed:2019mrd}%
  \BibitemOpen
  \bibfield  {author} {\bibinfo {author} {\bibfnamefont {N.}~\bibnamefont {Arkani-Hamed}}, \bibinfo {author} {\bibfnamefont {S.}~\bibnamefont {He}}, \ and\ \bibinfo {author} {\bibfnamefont {T.}~\bibnamefont {Lam}},\ }\href {\doibase 10.1007/JHEP02(2021)069} {\bibfield  {journal} {\bibinfo  {journal} {JHEP}\ }\textbf {\bibinfo {volume} {02}},\ \bibinfo {pages} {069} (\bibinfo {year} {2021})},\ \Eprint {http://arxiv.org/abs/1912.08707} {arXiv:1912.08707 [hep-th]} \BibitemShut {NoStop}%
\bibitem [{\citenamefont {Gon\c{c}alves}\ \emph {et~al.}(2025)\citenamefont {Gon\c{c}alves}, \citenamefont {Nocchi},\ and\ \citenamefont {Zhou}}]{Goncalves:2025jcg}%
  \BibitemOpen
  \bibfield  {author} {\bibinfo {author} {\bibfnamefont {V.}~\bibnamefont {Gon\c{c}alves}}, \bibinfo {author} {\bibfnamefont {M.}~\bibnamefont {Nocchi}}, \ and\ \bibinfo {author} {\bibfnamefont {X.}~\bibnamefont {Zhou}},\ }\href {\doibase 10.1007/JHEP06(2025)173} {\bibfield  {journal} {\bibinfo  {journal} {JHEP}\ }\textbf {\bibinfo {volume} {06}},\ \bibinfo {pages} {173} (\bibinfo {year} {2025})},\ \Eprint {http://arxiv.org/abs/2502.10269} {arXiv:2502.10269 [hep-th]} \BibitemShut {NoStop}%
\bibitem [{\citenamefont {de~Rham}\ \emph {et~al.}(2011)\citenamefont {de~Rham}, \citenamefont {Gabadadze},\ and\ \citenamefont {Tolley}}]{deRham:2010kj}%
  \BibitemOpen
  \bibfield  {author} {\bibinfo {author} {\bibfnamefont {C.}~\bibnamefont {de~Rham}}, \bibinfo {author} {\bibfnamefont {G.}~\bibnamefont {Gabadadze}}, \ and\ \bibinfo {author} {\bibfnamefont {A.~J.}\ \bibnamefont {Tolley}},\ }\href {\doibase 10.1103/PhysRevLett.106.231101} {\bibfield  {journal} {\bibinfo  {journal} {Phys. Rev. Lett.}\ }\textbf {\bibinfo {volume} {106}},\ \bibinfo {pages} {231101} (\bibinfo {year} {2011})},\ \Eprint {http://arxiv.org/abs/1011.1232} {arXiv:1011.1232 [hep-th]} \BibitemShut {NoStop}%
\bibitem [{\citenamefont {Momeni}\ \emph {et~al.}(2020)\citenamefont {Momeni}, \citenamefont {Rumbutis},\ and\ \citenamefont {Tolley}}]{Momeni:2020vvr}%
  \BibitemOpen
  \bibfield  {author} {\bibinfo {author} {\bibfnamefont {A.}~\bibnamefont {Momeni}}, \bibinfo {author} {\bibfnamefont {J.}~\bibnamefont {Rumbutis}}, \ and\ \bibinfo {author} {\bibfnamefont {A.~J.}\ \bibnamefont {Tolley}},\ }\href {\doibase 10.1007/JHEP12(2020)030} {\bibfield  {journal} {\bibinfo  {journal} {JHEP}\ }\textbf {\bibinfo {volume} {12}},\ \bibinfo {pages} {030} (\bibinfo {year} {2020})},\ \Eprint {http://arxiv.org/abs/2004.07853} {arXiv:2004.07853 [hep-th]} \BibitemShut {NoStop}%
\bibitem [{\citenamefont {Costa}\ \emph {et~al.}(2014)\citenamefont {Costa}, \citenamefont {Goncalves},\ and\ \citenamefont {Penedones}}]{Costa:2014kfa}%
  \BibitemOpen
  \bibfield  {author} {\bibinfo {author} {\bibfnamefont {M.~S.}\ \bibnamefont {Costa}}, \bibinfo {author} {\bibfnamefont {V.}~\bibnamefont {Goncalves}}, \ and\ \bibinfo {author} {\bibfnamefont {J.}~\bibnamefont {Penedones}},\ }\href {\doibase 10.1007/JHEP09(2014)064} {\bibfield  {journal} {\bibinfo  {journal} {JHEP}\ }\textbf {\bibinfo {volume} {09}},\ \bibinfo {pages} {064} (\bibinfo {year} {2014})},\ \Eprint {http://arxiv.org/abs/1404.5625} {arXiv:1404.5625 [hep-th]} \BibitemShut {NoStop}%
\bibitem [{\citenamefont {Li}\ and\ \citenamefont {Mei}(2023)}]{Li:2023azu}%
  \BibitemOpen
  \bibfield  {author} {\bibinfo {author} {\bibfnamefont {Y.-Z.}\ \bibnamefont {Li}}\ and\ \bibinfo {author} {\bibfnamefont {J.}~\bibnamefont {Mei}},\ }\href {\doibase 10.1007/JHEP07(2023)156} {\bibfield  {journal} {\bibinfo  {journal} {JHEP}\ }\textbf {\bibinfo {volume} {07}},\ \bibinfo {pages} {156} (\bibinfo {year} {2023})},\ \Eprint {http://arxiv.org/abs/2304.12757} {arXiv:2304.12757 [hep-th]} \BibitemShut {NoStop}%
\bibitem [{Note1()}]{Note1}%
  \BibitemOpen
  \bibinfo {note} {The total number of such graphs is nicely given in \protect \url {https://oeis.org/A001002}.}\BibitemShut {Stop}%
\bibitem [{Note2()}]{Note2}%
  \BibitemOpen
  \bibinfo {note} {In general, $s_{I}\equiv -2-\DOTSB \sum@ \slimits@ _{a,b\in I}{\protect \bf kk}_{ab}$ for an arbitrary subset $I$ of indices.}\BibitemShut {Stop}%
\bibitem [{Note3()}]{Note3}%
  \BibitemOpen
  \bibinfo {note} {For each vertex we have three possible levels, $m_v^{(1,2,3)}$, which are given by descendant levels on the corresponding propagators.}\BibitemShut {Stop}%
\bibitem [{Note4()}]{Note4}%
  \BibitemOpen
  \bibinfo {note} {For $n=4,5$, every propagator connects to a cubic vertex with two external legs, so the contraction $V \bullet V$ can be equivalently performed with $\eta ^{\mu \nu }$. However, this is no longer true for some graphs starting from $n=6$.}\BibitemShut {Stop}%
\bibitem [{\citenamefont {Bern}\ \emph {et~al.}(2008)\citenamefont {Bern}, \citenamefont {Carrasco},\ and\ \citenamefont {Johansson}}]{Bern:2008qj}%
  \BibitemOpen
  \bibfield  {author} {\bibinfo {author} {\bibfnamefont {Z.}~\bibnamefont {Bern}}, \bibinfo {author} {\bibfnamefont {J.~J.~M.}\ \bibnamefont {Carrasco}}, \ and\ \bibinfo {author} {\bibfnamefont {H.}~\bibnamefont {Johansson}},\ }\href {\doibase 10.1103/PhysRevD.78.085011} {\bibfield  {journal} {\bibinfo  {journal} {Phys. Rev. D}\ }\textbf {\bibinfo {volume} {78}},\ \bibinfo {pages} {085011} (\bibinfo {year} {2008})},\ \Eprint {http://arxiv.org/abs/0805.3993} {arXiv:0805.3993 [hep-ph]} \BibitemShut {NoStop}%
\bibitem [{\citenamefont {Bern}\ \emph {et~al.}(2010)\citenamefont {Bern}, \citenamefont {Carrasco},\ and\ \citenamefont {Johansson}}]{Bern:2010ue}%
  \BibitemOpen
  \bibfield  {author} {\bibinfo {author} {\bibfnamefont {Z.}~\bibnamefont {Bern}}, \bibinfo {author} {\bibfnamefont {J.~J.~M.}\ \bibnamefont {Carrasco}}, \ and\ \bibinfo {author} {\bibfnamefont {H.}~\bibnamefont {Johansson}},\ }\href {\doibase 10.1103/PhysRevLett.105.061602} {\bibfield  {journal} {\bibinfo  {journal} {Phys. Rev. Lett.}\ }\textbf {\bibinfo {volume} {105}},\ \bibinfo {pages} {061602} (\bibinfo {year} {2010})},\ \Eprint {http://arxiv.org/abs/1004.0476} {arXiv:1004.0476 [hep-th]} \BibitemShut {NoStop}%
\bibitem [{\citenamefont {Dong}\ \emph {et~al.}(2022)\citenamefont {Dong}, \citenamefont {He},\ and\ \citenamefont {Hou}}]{Dong:2021qai}%
  \BibitemOpen
  \bibfield  {author} {\bibinfo {author} {\bibfnamefont {J.}~\bibnamefont {Dong}}, \bibinfo {author} {\bibfnamefont {S.}~\bibnamefont {He}}, \ and\ \bibinfo {author} {\bibfnamefont {L.}~\bibnamefont {Hou}},\ }\href {\doibase 10.1103/PhysRevD.105.105007} {\bibfield  {journal} {\bibinfo  {journal} {Phys. Rev. D}\ }\textbf {\bibinfo {volume} {105}},\ \bibinfo {pages} {105007} (\bibinfo {year} {2022})},\ \Eprint {http://arxiv.org/abs/2111.10525} {arXiv:2111.10525 [hep-th]} \BibitemShut {NoStop}%
\bibitem [{\citenamefont {Berends}\ and\ \citenamefont {Giele}(1988)}]{Berends:1987me}%
  \BibitemOpen
  \bibfield  {author} {\bibinfo {author} {\bibfnamefont {F.~A.}\ \bibnamefont {Berends}}\ and\ \bibinfo {author} {\bibfnamefont {W.~T.}\ \bibnamefont {Giele}},\ }\href {\doibase 10.1016/0550-3213(88)90442-7} {\bibfield  {journal} {\bibinfo  {journal} {Nucl. Phys. B}\ }\textbf {\bibinfo {volume} {306}},\ \bibinfo {pages} {759} (\bibinfo {year} {1988})}\BibitemShut {NoStop}%
\bibitem [{\citenamefont {Britto}\ \emph {et~al.}(2005)\citenamefont {Britto}, \citenamefont {Cachazo}, \citenamefont {Feng},\ and\ \citenamefont {Witten}}]{Britto:2005fq}%
  \BibitemOpen
  \bibfield  {author} {\bibinfo {author} {\bibfnamefont {R.}~\bibnamefont {Britto}}, \bibinfo {author} {\bibfnamefont {F.}~\bibnamefont {Cachazo}}, \bibinfo {author} {\bibfnamefont {B.}~\bibnamefont {Feng}}, \ and\ \bibinfo {author} {\bibfnamefont {E.}~\bibnamefont {Witten}},\ }\href {\doibase 10.1103/PhysRevLett.94.181602} {\bibfield  {journal} {\bibinfo  {journal} {Phys. Rev. Lett.}\ }\textbf {\bibinfo {volume} {94}},\ \bibinfo {pages} {181602} (\bibinfo {year} {2005})},\ \Eprint {http://arxiv.org/abs/hep-th/0501052} {arXiv:hep-th/0501052} \BibitemShut {NoStop}%
\bibitem [{\citenamefont {Aprile}\ and\ \citenamefont {Vieira}(2020)}]{Aprile:2020luw}%
  \BibitemOpen
  \bibfield  {author} {\bibinfo {author} {\bibfnamefont {F.}~\bibnamefont {Aprile}}\ and\ \bibinfo {author} {\bibfnamefont {P.}~\bibnamefont {Vieira}},\ }\href {\doibase 10.1007/JHEP12(2020)206} {\bibfield  {journal} {\bibinfo  {journal} {JHEP}\ }\textbf {\bibinfo {volume} {12}},\ \bibinfo {pages} {206} (\bibinfo {year} {2020})},\ \Eprint {http://arxiv.org/abs/2007.09176} {arXiv:2007.09176 [hep-th]} \BibitemShut {NoStop}%
\bibitem [{\citenamefont {Abl}\ \emph {et~al.}(2021)\citenamefont {Abl}, \citenamefont {Heslop},\ and\ \citenamefont {Lipstein}}]{Abl:2020dbx}%
  \BibitemOpen
  \bibfield  {author} {\bibinfo {author} {\bibfnamefont {T.}~\bibnamefont {Abl}}, \bibinfo {author} {\bibfnamefont {P.}~\bibnamefont {Heslop}}, \ and\ \bibinfo {author} {\bibfnamefont {A.~E.}\ \bibnamefont {Lipstein}},\ }\href {\doibase 10.1007/JHEP04(2021)237} {\bibfield  {journal} {\bibinfo  {journal} {JHEP}\ }\textbf {\bibinfo {volume} {04}},\ \bibinfo {pages} {237} (\bibinfo {year} {2021})},\ \Eprint {http://arxiv.org/abs/2012.12091} {arXiv:2012.12091 [hep-th]} \BibitemShut {NoStop}%
\bibitem [{\citenamefont {Aprile}\ \emph {et~al.}(2021)\citenamefont {Aprile}, \citenamefont {Drummond}, \citenamefont {Paul},\ and\ \citenamefont {Santagata}}]{Aprile:2020mus}%
  \BibitemOpen
  \bibfield  {author} {\bibinfo {author} {\bibfnamefont {F.}~\bibnamefont {Aprile}}, \bibinfo {author} {\bibfnamefont {J.~M.}\ \bibnamefont {Drummond}}, \bibinfo {author} {\bibfnamefont {H.}~\bibnamefont {Paul}}, \ and\ \bibinfo {author} {\bibfnamefont {M.}~\bibnamefont {Santagata}},\ }\href {\doibase 10.1007/JHEP11(2021)109} {\bibfield  {journal} {\bibinfo  {journal} {JHEP}\ }\textbf {\bibinfo {volume} {11}},\ \bibinfo {pages} {109} (\bibinfo {year} {2021})},\ \Eprint {http://arxiv.org/abs/2012.12092} {arXiv:2012.12092 [hep-th]} \BibitemShut {NoStop}%
\bibitem [{\citenamefont {Aprile}\ \emph {et~al.}(2023)\citenamefont {Aprile}, \citenamefont {Drummond}, \citenamefont {Glew},\ and\ \citenamefont {Santagata}}]{Aprile:2022tzr}%
  \BibitemOpen
  \bibfield  {author} {\bibinfo {author} {\bibfnamefont {F.}~\bibnamefont {Aprile}}, \bibinfo {author} {\bibfnamefont {J.~M.}\ \bibnamefont {Drummond}}, \bibinfo {author} {\bibfnamefont {R.}~\bibnamefont {Glew}}, \ and\ \bibinfo {author} {\bibfnamefont {M.}~\bibnamefont {Santagata}},\ }\href {\doibase 10.1007/JHEP02(2023)087} {\bibfield  {journal} {\bibinfo  {journal} {JHEP}\ }\textbf {\bibinfo {volume} {02}},\ \bibinfo {pages} {087} (\bibinfo {year} {2023})},\ \Eprint {http://arxiv.org/abs/2207.13084} {arXiv:2207.13084 [hep-th]} \BibitemShut {NoStop}%
\bibitem [{\citenamefont {Alday}\ \emph {et~al.}(2022{\natexlab{c}})\citenamefont {Alday}, \citenamefont {Hansen},\ and\ \citenamefont {Silva}}]{Alday:2022uxp}%
  \BibitemOpen
  \bibfield  {author} {\bibinfo {author} {\bibfnamefont {L.~F.}\ \bibnamefont {Alday}}, \bibinfo {author} {\bibfnamefont {T.}~\bibnamefont {Hansen}}, \ and\ \bibinfo {author} {\bibfnamefont {J.~A.}\ \bibnamefont {Silva}},\ }\href {\doibase 10.1007/JHEP10(2022)036} {\bibfield  {journal} {\bibinfo  {journal} {JHEP}\ }\textbf {\bibinfo {volume} {10}},\ \bibinfo {pages} {036} (\bibinfo {year} {2022}{\natexlab{c}})},\ \Eprint {http://arxiv.org/abs/2204.07542} {arXiv:2204.07542 [hep-th]} \BibitemShut {NoStop}%
\bibitem [{\citenamefont {Alday}\ \emph {et~al.}(2022{\natexlab{d}})\citenamefont {Alday}, \citenamefont {Hansen},\ and\ \citenamefont {Silva}}]{Alday:2022xwz}%
  \BibitemOpen
  \bibfield  {author} {\bibinfo {author} {\bibfnamefont {L.~F.}\ \bibnamefont {Alday}}, \bibinfo {author} {\bibfnamefont {T.}~\bibnamefont {Hansen}}, \ and\ \bibinfo {author} {\bibfnamefont {J.~A.}\ \bibnamefont {Silva}},\ }\href {\doibase 10.1007/JHEP12(2022)010} {\bibfield  {journal} {\bibinfo  {journal} {JHEP}\ }\textbf {\bibinfo {volume} {12}},\ \bibinfo {pages} {010} (\bibinfo {year} {2022}{\natexlab{d}})},\ \Eprint {http://arxiv.org/abs/2209.06223} {arXiv:2209.06223 [hep-th]} \BibitemShut {NoStop}%
\bibitem [{\citenamefont {Alday}\ \emph {et~al.}(2023)\citenamefont {Alday}, \citenamefont {Hansen},\ and\ \citenamefont {Silva}}]{Alday:2023jdk}%
  \BibitemOpen
  \bibfield  {author} {\bibinfo {author} {\bibfnamefont {L.~F.}\ \bibnamefont {Alday}}, \bibinfo {author} {\bibfnamefont {T.}~\bibnamefont {Hansen}}, \ and\ \bibinfo {author} {\bibfnamefont {J.~A.}\ \bibnamefont {Silva}},\ }\href {\doibase 10.1103/PhysRevLett.131.161603} {\bibfield  {journal} {\bibinfo  {journal} {Phys. Rev. Lett.}\ }\textbf {\bibinfo {volume} {131}},\ \bibinfo {pages} {161603} (\bibinfo {year} {2023})},\ \Eprint {http://arxiv.org/abs/2305.03593} {arXiv:2305.03593 [hep-th]} \BibitemShut {NoStop}%
\bibitem [{\citenamefont {Alday}\ and\ \citenamefont {Hansen}(2023)}]{Alday:2023mvu}%
  \BibitemOpen
  \bibfield  {author} {\bibinfo {author} {\bibfnamefont {L.~F.}\ \bibnamefont {Alday}}\ and\ \bibinfo {author} {\bibfnamefont {T.}~\bibnamefont {Hansen}},\ }\href {\doibase 10.1007/JHEP10(2023)023} {\bibfield  {journal} {\bibinfo  {journal} {JHEP}\ }\textbf {\bibinfo {volume} {10}},\ \bibinfo {pages} {023} (\bibinfo {year} {2023})},\ \Eprint {http://arxiv.org/abs/2306.12786} {arXiv:2306.12786 [hep-th]} \BibitemShut {NoStop}%
\bibitem [{\citenamefont {Wang}\ \emph {et~al.}(2025)\citenamefont {Wang}, \citenamefont {Wu},\ and\ \citenamefont {Yuan}}]{Wang:2025pjo}%
  \BibitemOpen
  \bibfield  {author} {\bibinfo {author} {\bibfnamefont {B.}~\bibnamefont {Wang}}, \bibinfo {author} {\bibfnamefont {D.}~\bibnamefont {Wu}}, \ and\ \bibinfo {author} {\bibfnamefont {E.~Y.}\ \bibnamefont {Yuan}},\ }\href {\doibase 10.1103/v72s-rv7y} {\bibfield  {journal} {\bibinfo  {journal} {Phys. Rev. Lett.}\ }\textbf {\bibinfo {volume} {135}},\ \bibinfo {pages} {041603} (\bibinfo {year} {2025})},\ \Eprint {http://arxiv.org/abs/2503.01964} {arXiv:2503.01964 [hep-th]} \BibitemShut {NoStop}%
\bibitem [{\citenamefont {Loebbert}\ \emph {et~al.}(2018)\citenamefont {Loebbert}, \citenamefont {Mojaza},\ and\ \citenamefont {Plefka}}]{Loebbert:2018xce}%
  \BibitemOpen
  \bibfield  {author} {\bibinfo {author} {\bibfnamefont {F.}~\bibnamefont {Loebbert}}, \bibinfo {author} {\bibfnamefont {M.}~\bibnamefont {Mojaza}}, \ and\ \bibinfo {author} {\bibfnamefont {J.}~\bibnamefont {Plefka}},\ }\href {\doibase 10.1007/JHEP05(2018)208} {\bibfield  {journal} {\bibinfo  {journal} {JHEP}\ }\textbf {\bibinfo {volume} {05}},\ \bibinfo {pages} {208} (\bibinfo {year} {2018})},\ \Eprint {http://arxiv.org/abs/1802.05999} {arXiv:1802.05999 [hep-th]} \BibitemShut {NoStop}%
\bibitem [{\citenamefont {Caron-Huot}\ and\ \citenamefont {Trinh}(2019)}]{Caron-Huot:2018kta}%
  \BibitemOpen
  \bibfield  {author} {\bibinfo {author} {\bibfnamefont {S.}~\bibnamefont {Caron-Huot}}\ and\ \bibinfo {author} {\bibfnamefont {A.-K.}\ \bibnamefont {Trinh}},\ }\href {\doibase 10.1007/JHEP01(2019)196} {\bibfield  {journal} {\bibinfo  {journal} {JHEP}\ }\textbf {\bibinfo {volume} {01}},\ \bibinfo {pages} {196} (\bibinfo {year} {2019})},\ \Eprint {http://arxiv.org/abs/1809.09173} {arXiv:1809.09173 [hep-th]} \BibitemShut {NoStop}%
\bibitem [{\citenamefont {Caron-Huot}\ and\ \citenamefont {Coronado}(2022)}]{Caron-Huot:2021usw}%
  \BibitemOpen
  \bibfield  {author} {\bibinfo {author} {\bibfnamefont {S.}~\bibnamefont {Caron-Huot}}\ and\ \bibinfo {author} {\bibfnamefont {F.}~\bibnamefont {Coronado}},\ }\href {\doibase 10.1007/JHEP03(2022)151} {\bibfield  {journal} {\bibinfo  {journal} {JHEP}\ }\textbf {\bibinfo {volume} {03}},\ \bibinfo {pages} {151} (\bibinfo {year} {2022})},\ \Eprint {http://arxiv.org/abs/2106.03892} {arXiv:2106.03892 [hep-th]} \BibitemShut {NoStop}%
\bibitem [{\citenamefont {Caron-Huot}\ \emph {et~al.}(2023)\citenamefont {Caron-Huot}, \citenamefont {Coronado},\ and\ \citenamefont {M{\"u}hlmann}}]{Caron-Huot:2023wdh}%
  \BibitemOpen
  \bibfield  {author} {\bibinfo {author} {\bibfnamefont {S.}~\bibnamefont {Caron-Huot}}, \bibinfo {author} {\bibfnamefont {F.}~\bibnamefont {Coronado}}, \ and\ \bibinfo {author} {\bibfnamefont {B.}~\bibnamefont {M{\"u}hlmann}},\ }\href {\doibase 10.1007/JHEP08(2023)008} {\bibfield  {journal} {\bibinfo  {journal} {JHEP}\ }\textbf {\bibinfo {volume} {08}},\ \bibinfo {pages} {008} (\bibinfo {year} {2023})},\ \Eprint {http://arxiv.org/abs/2304.12341} {arXiv:2304.12341 [hep-th]} \BibitemShut {NoStop}%
\bibitem [{\citenamefont {Huang}\ \emph {et~al.}(2025{\natexlab{b}})\citenamefont {Huang}, \citenamefont {Wang}, \citenamefont {Yuan},\ and\ \citenamefont {Zhang}}]{Huang:2024dxr}%
  \BibitemOpen
  \bibfield  {author} {\bibinfo {author} {\bibfnamefont {Z.}~\bibnamefont {Huang}}, \bibinfo {author} {\bibfnamefont {B.}~\bibnamefont {Wang}}, \bibinfo {author} {\bibfnamefont {E.~Y.}\ \bibnamefont {Yuan}}, \ and\ \bibinfo {author} {\bibfnamefont {J.}~\bibnamefont {Zhang}},\ }\href {\doibase 10.1103/PhysRevLett.134.161601} {\bibfield  {journal} {\bibinfo  {journal} {Phys. Rev. Lett.}\ }\textbf {\bibinfo {volume} {134}},\ \bibinfo {pages} {161601} (\bibinfo {year} {2025}{\natexlab{b}})},\ \Eprint {http://arxiv.org/abs/2408.12260} {arXiv:2408.12260 [hep-th]} \BibitemShut {NoStop}%
\bibitem [{\citenamefont {Fernandes}\ \emph {et~al.}(2025)\citenamefont {Fernandes}, \citenamefont {Goncalves}, \citenamefont {Huang}, \citenamefont {Tang}, \citenamefont {Vilas~Boas},\ and\ \citenamefont {Yuan}}]{Fernandes:2025eqe}%
  \BibitemOpen
  \bibfield  {author} {\bibinfo {author} {\bibfnamefont {B.}~\bibnamefont {Fernandes}}, \bibinfo {author} {\bibfnamefont {V.}~\bibnamefont {Goncalves}}, \bibinfo {author} {\bibfnamefont {Z.}~\bibnamefont {Huang}}, \bibinfo {author} {\bibfnamefont {Y.}~\bibnamefont {Tang}}, \bibinfo {author} {\bibfnamefont {J.}~\bibnamefont {Vilas~Boas}}, \ and\ \bibinfo {author} {\bibfnamefont {E.~Y.}\ \bibnamefont {Yuan}},\ }\href@noop {} {\  (\bibinfo {year} {2025})},\ \Eprint {http://arxiv.org/abs/2507.14124} {arXiv:2507.14124 [hep-th]} \BibitemShut {NoStop}%
\bibitem [{\citenamefont {Sonine}(1880)}]{sonine_recherches_1880}%
  \BibitemOpen
  \bibfield  {author} {\bibinfo {author} {\bibfnamefont {N.}~\bibnamefont {Sonine}},\ }\href {\doibase 10.1007/BF01459227} {\bibfield  {journal} {\bibinfo  {journal} {Mathematische Annalen}\ }\textbf {\bibinfo {volume} {16}},\ \bibinfo {pages} {1} (\bibinfo {year} {1880})}\BibitemShut {NoStop}%
\end{thebibliography}%

\appendix

\onecolumngrid

\section{A: Review of super-Yang-Mills and supergravity in AdS\texorpdfstring{$_5$}{5} and results of mixed spin amplitudes}
\label{appendixA}
\subsection{Mellin amplitudes for super-Yang-Mills on \texorpdfstring{$\mathrm{AdS}_5 \times S^3$}{AdS5S3}}
We consider the $N$-point scalar amplitudes of the lowest KK modes in bulk $\mathcal{N}=1$ sYM on AdS$_5\times S^3$, or equivalently the connected correlators of half-BPS operators $\mathcal{O}^a(x,v)$ with conformal dimension $\Delta=2$ in boundary $\mathcal{N}=2$ SCFT:
\begin{equation}
    G_{N,\,\mathrm{scalar}}^{a_1\cdots a_{N}}=\langle\mathcal{O}^{a_1}(x_1,v_1)\cdots\mathcal{O}^{a_{N}}(x_{N},v_{N})\rangle\,.
\end{equation}
Here, $\mathcal{O}^a(x,v)=\mathcal{O}_2^{a;\alpha_1\alpha_2}(x)\,v^{\beta_1}v^{\beta_2}\,\epsilon_{\alpha_1\beta_1}\epsilon_{\alpha_2\beta_2}$, where $a_i$ denotes adjoint indices of $G_F$, a gauge symmetry in the bulk AdS that is dual to a global symmetry in the boundary CFT, and $ v^\beta $'s are auxiliary $\mathrm{SU}(2)_R$ R-symmetry spinors. To capture the vector exchanges, we also introduce the bulk ``gluon'' $A^a_\mu$, which is dual to the boundary $G_F$-Noether current $\mathcal{J}_\mu^a$ in the same supermultiplet as $\mathcal{O}^a$, and define the single-gluon correlators:
\begin{equation}
    G_{N,\,\text{single-gluon}}^{a_1\cdots a_{N}}=\langle\mathcal{O}^{a_1}(x_1)\cdots \mathcal{O}^{a_{N-1}}(x_{N-1}) \mathcal{J}^{a_{N}}(x_{N},z)\rangle,
\end{equation}
where $\mathcal{J}^{a}(x,z)$ is $\mathcal{J}_\mu^a(x)$ contracted with polarization $z^\mu$ and is an $\mathrm{SU}(2)_R$ singlet. At tree level, the Yang-Mills-scalar theory comprised of $\mathcal{O}^a(x,v)$ and $\mathcal{J}_\mu^a$ forms a closed sub-sector that decouples from higher KK modes and the gravitons, making it an ideal arena for the scaffolding formalism.

Analogous to the flat-space formalism, we perform the color decomposition for tree-level amplitudes in AdS~\cite{Cao:2023cwa}:
\begin{equation}
    G_N^{a_1\cdots a_N}=\sum_{\mathclap{\sigma\in S_{N-1}}} {\rm tr}(T^{a_1}T^{a_{\sigma(2)}}\cdots T^{a_{\sigma(N)}}) G_{1\sigma}\,,
\end{equation}
where the sum runs over permutations $\sigma$ of the indices $\{2,\dots, N\}$. In the main text, we focus on the partial amplitude $G_{12\cdots N}$ (with $N=2n$), as any other color-ordered amplitude relates to it by relabeling. The Mellin amplitudes for scalar correlators are defined by:
\begin{equation}
    G_{12\cdots N}^{\mathrm{scalar}}=\int[{\rm d}\delta]\mathcal M_N^{(\mathrm{s})}(\delta_{ij},v_i)\prod_{i<j}\frac{\Gamma(\delta_{ij})}{(-2P_i\cdot P_j)^{\delta_{ij}}}, \tag*{(\ref{def:Mellin})}
\end{equation}
and for single-gluon correlators (with the $N$-th leg being the gluon), we define~\cite{Goncalves:2014rfa,Cao:2024bky}:
\begin{gather}\label{def:single-vector}
    G_{12\cdots N}^{\text{single-gluon}}=\int[{\rm d}\delta]\mathcal M_N^{(\mathrm{v})}\prod_{i<j}\frac{\Gamma(\delta_{ij})}{(-2P_i\cdot P_j)^{\delta_{ij}}}\,, \quad \mathcal M_N^{(\mathrm{v})}= \sum_{\ell=1}^{N-1}\mathcal M_N^{(\mathrm{v})\ell}(\delta_{ij},v_i)\,\texttt{zp[$\ell$]}\\[-2pt]
    \text{where}\quad\texttt{zp[$\ell$]}=\frac{(Z_N\cdot P_\ell)}{(-2P_\ell\cdot P_N)}\delta_{\ell N}\,,\quad\sum_{\ell=1}^{N-1}\delta_{\ell N}\mathcal M_N^{(\mathrm{v})\ell}=0,\nonumber
\end{gather}
where we use the embedding formalism~\cite{Goncalves:2014rfa}, where $(x_i-x_j)^2=-2P_i\cdot P_j$ and $Z_N\cdot P_\ell$ encodes the vector structure of $\mathcal J_\mu^a$. Note that the definition of $\mathcal M_N^{(\mathrm{v})}$ implicitly absorbs the Kronecker delta shift $\delta_i^\ell\delta_j^N$ in Gamma functions.

The Mellin amplitudes defined above generally carry some R-symmetry structures. For brevity of notations, we define $V_{i_1 i_2 \cdots i_r}\equiv\left\langle i_1 i_2\right\rangle\left\langle i_2 i_3\right\rangle \cdots\left\langle i_r i_1\right\rangle$, with $\langle i j\rangle:=v_i^\alpha v_j^\beta \epsilon_{\alpha \beta}$. Crucially, due to the R-charge neutrality of $\mathcal{J}_\mu^a$, the scaffolded $n$-gluon Mellin amplitudes must carry the ``$n$-trace'' R-structure $V_{12}V_{34}\cdots V_{2n-1,2n}$. This allows us to omit the explicit R-factor in the main text. For example, restoring this factor, the scaffolded 3-gluon amplitude is given by
\begin{equation}
  \mathcal{A}_{3,\; \mathrm{AdS}_5 \times S^3}^{(1,1,1)}=2\left(\mathbf{e e }_ { 1 2 } \mathbf{e k}_{31}+\mathbf{e e}_{23} \mathbf{e k}_{12}+\mathbf{e e}_{31} \mathbf{e k}_{23}\right)V_{12}V_{34}V_{56}\,,
\end{equation}
whereas mixed-spin amplitudes may involve more complex R-structures, such as:
\begin{equation}
    \mathcal{A}_{3,\; \mathrm{AdS}_5 \times S^3}^{(1,0,0)}=\frac{1}{2}{\bf ek}_{13} \, V_{12}(V_{34}V_{56}-V_{3456}) \,.
\end{equation}

\subsection{Mellin amplitudes for Supergravity on \texorpdfstring{$\mathrm{AdS}_5 \times S^5$}{AdS5S5}}
We consider the $N$-point scalar amplitudes of the lowest KK modes in bulk type IIB supergravity on $\mathrm{AdS}_5\times S^5$, which correspond to the connected correlators of half-BPS operators $\mathcal{O}(x,y)$ with conformal dimension $\Delta=2$ in the holographically dual $\mathcal{N}=4$ sYM:
\begin{equation}
    G_{N,\,\mathrm{scalar}}=\langle\mathcal{O}(x_1,y_1)\cdots\mathcal{O}(x_{N},y_{N})\rangle\,.
\end{equation}
Here, $\mathcal{O}(x,y) = \mathcal{O}_2^{IJ}(x)\,y_Iy_J$, where $\mathcal{O}_2^{IJ}$ transforms in the rank-2 symmetric traceless representation of $SO(6)_R$, and $y_I$'s are auxiliary $SO(6)_R$ R-symmetry polarizations satisfying $y^2=0$. To capture spinning exchanges, we also introduce the bulk ``graviphoton'' $A^{IJ}_\mu$ and the bulk ``graviton'' $h_{\mu\nu}$ belonging to the same supermultiplet, which are dual to the boundary R-symmetry current $J_\mu^{[IJ]}$ and the stress tensor $\mathcal{T}^{\mu\nu}$ (R-singlet), respectively.

The Mellin amplitude $\mathcal M_N^{(\mathrm{s})}(\delta_{ij},y_i)$ for $G_{N,\,\mathrm{scalar}}$ is defined via the same representation as in~\eqref{def:Mellin}. The extraction of spinning amplitudes proceeds analogously to the supergluon case, but with distinct R-symmetry structures. Specifically, since $\mathcal{T}^{\mu\nu}$ is R-neutral, the scaffolded $n$-graviton amplitudes carry the ``$n$-trace'' R-structure $y_{12}^2y_{34}^2\cdots y_{2n-1,2n}^2$ (with $y_{ij}^2\equiv y_i\cdot y_j$). The graviphoton amplitudes, however, involve non-trivial R-structures. For example, extracting from the six-scalar results~\cite{Goncalves:2025jcg} using the projectors in Table~\ref{tab:spin2_projectors}, the ``three-photon'' amplitude reads:

\begin{equation}
    \mathcal{A}_{3,\;\mathrm{AdS}_5 \times S^5}^{(1,1,1)}=\frac{8}{3}\left(\mathbf{e e }_ { 1 2 } \mathbf{e k}_{31}+\mathbf{e e}_{23} \mathbf{e k}_{12}+\mathbf{e e}_{31} \mathbf{e k}_{23}\right)y_{12} y_{34} y_{56} \sum_{(i_a\,j_a) \neq (12),(34),(56)} (-)^{\sigma(i_a j_a)}y_{i_1, j_1} y_{i_2, j_2} y_{i_3, j_3}
\end{equation}
where the sum runs over the eight pairings of $\{1,2,\dots, 6\}$ (denoted as $(i_1 j_1), (i_2 j_2), (i_3 j_3)$) that do not contain $(12), (34)$, or $(56)$, with $(-)^{\sigma(i_a j_a)}$ denoting the sign of the permutation $(i_1j_1i_2j_2i_3j_3)$. Explicitly, it reads
$y_{16} (y_{24} y_{35}-y_{23} y_{45}) +y_{15} (y_{23} y_{46}- y_{24} y_{36})+y_{14} (y_{25} y_{36}-y_{26} y_{35})+y_{13}(y_{26} y_{45}-y_{25} y_{46})$. The kinematic part agrees precisely with the three-gluon amplitude $\mathcal{A}_{3}^{(1,1,1)}$ in the main text, which is antisymmetric under the swap $i \leftrightarrow j$. The Bose symmetry is recovered from that the R-symmetry factor is also antisymmetric under the simultaneous swaps $ 2i-1 \leftrightarrow 2j-1 $ and $ 2i  \leftrightarrow 2j$ for the $2n$-gon variables (similar to a structure constant $f_{a,b,c}$). 

We also extract the graviton-photon-photon amplitude as
\begin{equation}
    \mathcal{A}_{3,\;\mathrm{AdS}_5 \times S^5}^{(2,1,1)}=\left[4\; {\bf ek}_{13}({\bf ee}_{13}{\bf ek}_{21}+{\bf ee}_{23}{\bf ek}_{13}+{\bf ee}_{12}{\bf ek}_{32})-\frac{4}{3}{\bf ee}_{12}{\bf ee}_{13}\right] y_{12}^2 y_{34} y_{56}\left(y_{36} y_{45}-y_{35} y_{46}\right).
\end{equation}
We observe that the first term in the kinematic part is proportional to $\mathcal{A}_{3,\;\mathrm{AdS}_5 \times S^3}^{\rm (1,1,1)}\times\mathcal{A}_{3,\;\mathrm{AdS}_5 \times S^3}^{(1,0,0)}$, while the second term ${\bf ee}_{12}{\bf ee}_{13}$ remains. Bose symmetry is satisfied by the kinematic and R-symmetry factors individually.

The graviton-scalar-scalar amplitude reads
\begin{equation}
     \mathcal{A}_{3,\;\mathrm{AdS}_5 \times S^5}^{(2,0,0)}=\frac{1}{12} {\bf ek}_{13}^2 y_{12}^2 y_{34} y_{56} \left(8 y_{36} y_{45}+8 y_{35} y_{46}+y_{34} y_{56}\right).
\end{equation}
Here, the kinematic part is simply the square of $\mathcal{A}_{3,\;\mathrm{AdS}_5 \times S^3}^{(1,0,0)}$.

Finally, we extract three graviton amplitude as
\begin{equation}
\begin{aligned}
    \mathcal{A}_{3,\;\mathrm{AdS}_5 \times S^5}^{(2,2,2)}=& \left(4 \left ({\bf ee}_{12} {\bf ek}_{31}+{\bf ee}_{23} {\bf ek}_{12} +{\bf ee}_{31} {\bf ek}_{23} \right)^2 - 8\,{\bf ee}_{12} {\bf ee}_{23} {\bf ee}_{31}\right)\,y_{12}^2y_{34}^2y_{56}^2.
\end{aligned}
\label{spin2EKcc}
\end{equation}

\section{B: Details on projectors and factorizations}\label{appendixB}

In this appendix, we give the factorization for spin-1 amplitudes and derive the spin-1 projector. We work in $\mathcal{N}=1$ sYM on $\mathrm{AdS}_5\times S^3$ for concreteness, but the results also apply to any Yang-Mills-scalar theory in $\mathrm{AdS}_{d+1}$ background. The factorization of the $N$-scalar Mellin amplitudes (with $N=2n$ in the main text) on gluon-exchanges reads~\cite{Goncalves:2014rfa}
\begin{equation}\label{eq:fac_spin_1}
    \overset{(\mathrm{v})}{\underset{\mathbf{X}_{1,k}=m}{\operatorname{Res}}}\mathcal{M}_{N}^{(\mathrm{s})}=\mathcal{N}_{\mathrm{v}}^{(m)}\sum_{a=1}^{k-1}\sum_{i=k}^N\delta_{ai}\mathcal{M}_{1\cdots(k-1)I}^{(\mathrm{v})(m)a}\mathcal{M}_{k\cdots NI}^{(\mathrm{v})(m)i},
\end{equation}
where $\mathcal{N}_{\mathrm{v}}^{(m)}=-\frac{(\Delta +J-1) \Gamma(\Delta -1) m!}{(-2)^{J}\left(\Delta-\frac{d}{2} +1\right)_m}$, which is $\frac{3}{2(1+m)}$ in our setup. The shifted amplitudes $\mathcal{M}_{1\cdots (k-1)I}^{(\mathrm{v})(m)a}$ are defined by
\begin{equation}\label{eq:partial_amplitude}
    \mathcal{M}_{1\cdots(k-1)I}^{(\mathrm{v})(m)a}=\sum_{\substack{n_{ij}\geqslant0\\\sum n_{ij}=m}}\mathcal{M}_{1\cdots(k-1)I}^{(\mathrm{v})a}(\delta_{ij}+n_{ij})\prod_{1\leqslant i<j<k}\frac{(\delta_{ij})_{n_{ij}}}{n_{ij}!}.
\end{equation}

Consider the case with $k=N-1$, in which $\mathcal{M}_{N-1,N,I}^{(\mathrm{v})}=\frac{\mathrm{i}}{\sqrt{6}}V_{N-1,N}(\texttt{zp[$N-1$]}-\texttt{zp[$N$]})$~\cite{Cao:2023cwa}. The residue on $\mathbf{X}_{1,N-1}=0$ reads
\begin{equation}\label{eq:residue_in_difference_delta}
\begin{aligned}
    \overset{(\mathrm{v})}{\underset{\mathbf{X}_{1,N-1}=0}{\operatorname{Res}}}\mathcal{M}_{N}^{(\mathrm s)}=\mathrm{i}\mathcal{N}_{\mathrm{v}}^{(0)}\frac{V_{N-1,N}}{\sqrt{6}}\sum_{a=1}^{N-2}\mathcal{M}_{N-1}^{(\mathrm{v})a}(\delta_{a,N-1}-\delta_{a,N}).
\end{aligned}
\end{equation}
Converting $\delta$ to $\mathbf{X}$, following the derivation as in~\cite{Cao:2023cwa,Cao:2024bky}, we have:
\begin{equation}\label{eq:residue_difference_X}
    \mathcal{M}_{N-1}^{(\mathrm{v})a}-\mathcal{M}_{N-1}^{(\mathrm{v})a-1}=\frac{\mathrm{i}\sqrt{3/2}}{\mathcal{N}_{v}^{(0)}V_{N-1,N}}\frac{\partial}{\partial \mathbf{X}_{a,N}}\left( \overset{(\mathrm{v})}{\underset{\mathbf{X}_{1,N-1}=0}{\mathrm{Res}}}\mathcal{M}_{N}^{(\mathrm{s})} \right).
\end{equation}
Solving \eqref{eq:residue_difference_X} together with the constrain $\sum_{a=1}^{N-2}{\delta_{a,N-1}\mathcal{M}_{N-1}^{(\mathrm{v})a}}=0$~\eqref{def:single-vector}, the single-gluon amplitudes $\mathcal{M}_{N-1}^{(\mathrm{v})a}$ can be extracted from the scalar amplitudes as
\begin{equation}\label{eq:to-single-vector-M}
    \mathcal{M}_{N-1}^{(\mathrm{v})a}=\frac{-\mathrm{i}}{\sqrt{6}V_{N-1,N}} \sum_{i=2}^{N-2}\left[\mathbf{X}_{i,N-1}-\mathbf{X}_{i,1}-\operatorname{sgn}(a-i)\right]\frac{\partial}{\partial\mathbf{X}_{i,N}}\overset{(\mathrm{v})}{\underset{\mathbf{X}_{1,N-1}=0}{\operatorname{Res}}}\mathcal{M}_N^{(\mathrm{s})}\,,\quad \operatorname{sgn}(x) =\begin{cases}
        +1 & \text{if }\;x\geqslant 0\\ -1 & \text{if }\;x<0
    \end{cases}.
\end{equation}
Substituting $N=2n$, \eqref{eq:to-single-vector-M} combined with~\eqref{eq:fac_spin_1} gives the factorization of scaffolded $n$-spinning amplitudes presented in~\eqref{eq:YMS-fac} of main text. We leave the detailed investigation of factorization in terms of the formal momenta ${\bf k}_i$ and polarizations ${\bf e}_i$ for future work.

We now proceed to verify that the operator $\mathcal{P}^{(1)}_i=\mathcal{P}^{({\bf e})}_i$ introduced in~\eqref{def:projector} correctly isolates the spin-1 contribution. Specifically, we demonstrate that (a) the projector is invariant on vector-exchange residues, and (b) it annihilates scalar-exchange residues. Regarding (a), we substitute~\eqref{eq:to-single-vector-M} into~\eqref{eq:residue_in_difference_delta} (with $N=2n$) and utilize the identity:
\begin{equation}\nonumber
\begin{aligned}
    &\textstyle\sum_{a=1}^{2n-2}(\delta_{a,2n-1}-\delta_{a,2n})[\mathbf{X}_{i,2n-1}-\mathbf{X}_{i,1}-\operatorname{sgn}(a-i)] \\
    &= -\textstyle\sum_{a=1}^{2n-2}(\delta_{a,2n-1}-\delta_{a,2n})\operatorname{sgn}(a-i)
    = 2\textstyle\left(\mathbf{X}_{i,2n}-\frac{\mathbf{X}_{i,1}+\mathbf{X}_{i,2n-1}-\mathbf{X}_{1,2n-1}+1}{2}\right),
\end{aligned}
\end{equation}
then we immediately recover the original residue with the projector emerging:
\begin{equation}\label{eq:projection_onv_res}
     \overset{(\mathrm{v})}{\underset{{\bf X}_{1,2n-1}=0}{\operatorname{Res}}}\mathcal{M}_{2n}^{(\mathrm{s})}=\sum_{i=2}^{2n-2} \left({\bf X}_{i,2n}-\frac{{\bf X}_{i,1}+\mathbf{X}_{i,2n-1}+1}{2}\right) \frac{\partial}{\partial{\bf X}_{i,2n}} \overset{(\mathrm{v})}{\underset{{\bf X}_{1,2n-1}=0}{\operatorname{Res}}}\mathcal{M}_{2n}^{(\mathrm{s})}.
\end{equation}
This confirms that the vector-exchange residue is invariant under $\mathcal{P}_{n}^{(1)}$. Regarding (b), note that the factorization on a scalar pole is given by~\cite{Goncalves:2014rfa,Cao:2023cwa}:
\begin{equation}\label{eq:scalar-fac}
    \overset{(\mathrm{s})}{\underset{{\bf X}_{1,2n-1}=0}{\operatorname{Res}}}\mathcal{M}_{2n}^{(\mathrm{s})} = \mathcal{N}_{\mathrm{s}}^{(m)} \texttt{glueR} \left(\mathcal{M}_{12\cdots(2n-2)I}^{(\mathrm{s})}\,\mathcal{M}_{(2n-1)(2n)I}^{(\mathrm{s})}\right)\,,\quad
    \texttt{glueR}\left(V_{i\cdots jI}\otimes V_{Ik\cdots l}\right)= V_{i\cdots jk\cdots l}-\frac12V_{i\cdots j}V_{k\cdots l}
\end{equation}
where $\mathcal{N}_{\mathrm{s}}^{(m)}=-1$ and $\texttt{glueR}$ is an operator acting purely on R-structure but not on kinematics. Since neither $\mathcal{M}_{12\cdots(2n-2)I}^{(\mathrm{s})}$ nor $\mathcal{M}_{(2n-1)(2n)I}^{(\mathrm{s})}$ depends on the cross-variables ${\bf X}_{i,2n}$ derived in the projector, the action of $\mathcal{P}_{n}^{(1)}$ vanishes. We therefore conclude that the projector isolates the spin-1 contribution:
\begin{equation}\label{eq:projector property}
    \mathcal{P}_{i}^{(1)}\left({\underset{{\bf X}_{2i-1,2i+1}=0}{\operatorname{Res}}}\mathcal{M}_{2n}^{(\mathrm{s})}\right)=\overset{(\mathrm{v})}{\underset{{\bf X}_{2i-1,2i+1}=0}{\operatorname{Res}}}\mathcal{M}_{2n}^{(\mathrm{s})}.
\end{equation}

The full $n$-gluon amplitude is then extracted by acting with projectors on all legs:
\begin{equation}
    \mathcal{A}_n^{\mathrm{YMS}}= \mathcal{P}_1^{(1)}\mathcal{P}_2^{(1)} \cdots \mathcal{P}_n^{(1)}{\underset{\mathbf X_{1,3}=0}{\operatorname{Res}}}\; {\underset{\mathbf X_{3,5}=0}{\operatorname{Res}}} \cdots \underset{\mathbf X_{2 n-1,2 n+1}=0}{\operatorname{Res}} \mathcal{M}_{2 n}^{(\mathrm{s})}\,.
\end{equation}
where the order of projectors does not matter as one can verify that the projectors commute, $\left[\mathcal{P}_{i}^{(1)},\mathcal{P}_{j}^{(1)}\right]=0$.

\section{C: Recursion relations for coefficients and the flat-space limit}\label{appendixC}
In this appendix, we extend our discussion on vertex functions $\alpha_{m_1\dots m_k}$. The general form is given as
\begin{equation}
    \alpha_{m_1\dots m_k} = \sum_{n_1=0}^{m_1} \cdots\sum_{n_k=0}^{m_k}  \Gamma\left(\sum_{j=1}^kn_j+k\right)\prod_{j=1}^k\frac{(-m_j)_{n_j}}{n_j!(n_j+1)!},
\end{equation}
for $d=4$ and $\Delta=3$. As we have mentioned in the main text, a crucial identity which guarantees flat-space limit is the following decomposition of $4$-vertex coefficient into the ``gluing'' of two $3$-vertex ones (since $\alpha_{m_1,\dots, m_4}$ is manifestly crossing symmetric, any way for decomposing it should be equivalent): 
\begin{equation}\label{eq57}
    \alpha_{m_1,m_2,m_3,m_4}=\sum_{m_I=0}^\infty\alpha_{m_1,m_2,m_I}(m_I+1)\alpha_{m_I,m_3,m_4}=\sum_{m_I=0}^\infty\alpha_{m_2,m_3,m_I}(m_I+1)\alpha_{m_I,m_4,m_1}.
\end{equation}
Note that from the definition of $\alpha_{m_1, m_2, m_I}$ {\it etc.}, the first summation must be truncated to $0\leqslant m_I \leqslant \min (m_1+m_2+1, m_3+m_4+1)$ (and similarly for the second summation). It is clear that given \eqref{eq57}, any Feynman diagram with $3$- and $4$-vertices must pick up the same overall coefficient. Very nicely, \eqref{eq57} is actually the first instance of a beautiful recursion relation, which generalizes to the case of decomposing an $n$-point vertex into $(k+1)$- and $(n-k+1)$-point vertices,
\begin{equation}
    \alpha_{m_1\dots m_n}=\sum_{m_I=0}^\infty\alpha_{m_1\dots m_km_I}(m_I+1)\alpha_{m_Im_{k+1}\dots m_{n}}.
\end{equation}
Such recursion relation then states that for any external descendant levels, $m_1, m_2,\dots, m_n$, any (sub-)diagram will pick up an overall coefficient which is exactly $\alpha_{m_1, m_2,\dots, m_n}$! What is needed here is the case with all external primary ones, which is why we find the overall coefficient to be $\alpha_{0,\dots, 0}=(n{-}1)!$ . 

The proof of this combinatorial identity is straightforward. To start with, using the integration definition of the Gamma function
$\Gamma(z)\equiv\int_0^\infty e^{-t}t^{z-1}\mathrm dt$
 and substitute $z=\sum_{j=1}^kn_j+k$, we have
\begin{equation} \alpha_{m_1\dots m_k} = \int_0^\infty e^{-t} t^{k-1} \prod_{j=1}^k\frac{L_{m_j}^{(1)}(t)}{m_j+1}\mathrm dt, 
\end{equation}
in which we have used $L_{m_j}^{(1)}(t)$ as generalized Laguerre functions ~\cite{sonine_recherches_1880} 
\begin{equation}
    L_n^{(\alpha)}(x)\equiv\sum_{i=0}^n(-1)^i\begin{pmatrix}
        n+\alpha\\ n-i
    \end{pmatrix}\frac{x^i}{i!}
\end{equation}
at $\alpha=1$. By imposing the orthogonality of Laguerre functions
\begin{equation}
    \sum_{m=0}^\infty\frac{L_m^{(1)}(x)L_m^{(1)}(y)}{m+1}=\frac{e^x}{x}\delta(x-y),
\end{equation}
 we have
\begin{equation}
\begin{aligned}
    \sum_{m_I=0}^\infty\alpha_{m_1\dots m_km_I}(m_I+1)\alpha_{m_Im_{k+1}\dots m_n}&=\iint_0^\infty{\mathrm dx\mathrm dy} e^{-x-y}x^{k}y^{n-k}\prod_{i=1}^k\frac{L_{m_i}^{(1)}(x)}{m_i+1}\prod_{j=k+1}^{n}\frac{L_{m_j}^{(1)}(y)}{m_j+1}\sum_{m_I=0}^\infty\frac{L_{m_I}^{(1)}(x)L_{m_I}^{(1)}(y)}{m_I+1}\\
    &=\int_0^\infty\mathrm{d}xe^{-x}x^{n-1}\prod_{i=1}^n\frac{L_{m_i}^{(1)}(x)}{m_i+1}=\alpha_{m_1\dots m_n}.
\end{aligned}
\end{equation}

Graphically, here we show explicitly how these coefficients up to $n=6$ conspire to give the correct flat-space limit with an overall $\alpha_{0,\dots, 0}=(n{-}1)!$ coefficient. We simply list seeds (dihedral inequivalent dual graphs) for $n= 4, 5$, and $6$: for each vertex, we denote its coefficient $\alpha_{\{m_v\}}$ in the corresponding sub-polygon, divided by propagators $\{I\}$, with the multiplicity of lines is given by the descendant level $m_I+1$ (see the following Tables~\ref{4pt seeds},~\ref{5pt seeds} and~\ref{6pt_seeds} for details).
\begin{table}[H]
\centering
\begin{tblr}
    {rowspec={|[1pt]Q[c,m]|Q[c,m]|[dotted]Q[c,m]|[1pt]},colspec={Q[c,m]|Q[c,m]|Q[c,m]}}
    Seeds  &Dual Graphs&$\sum_{\{m\}}\prod_{v\in V(g)}\prod_{I\in E(g)}(m_I+1)\alpha_{\{m_v\}}$\\
    \raisebox{-0.6cm}{\includegraphics[scale=0.7]{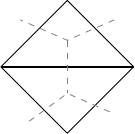}}&\raisebox{-0.6cm}{\includegraphics[scale=.7]{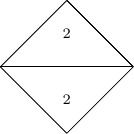}}\quad\raisebox{-0.6cm}{\includegraphics[scale=.7]{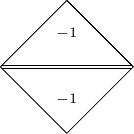}}&$4+2=3!$\\
    \raisebox{-0.6cm}{\includegraphics[scale=.7]{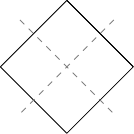}}&\raisebox{-0.6cm}{\includegraphics[scale=.7]{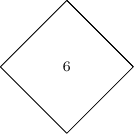}}&$6=3!$
\end{tblr}
\caption{All possible dual graphs for each seed and the total of vertex functions at 4-point level.}
\label{4pt seeds}
\end{table}
\begin{table}[H]
\centering
\begin{tblr}
    {rowspec={|[1pt]Q[c,m]|Q[c,m]|[dotted]Q[c,m]|[1pt]},colspec={Q[c,m]|Q[c,m]|Q[c,m]}}
    Seeds  &Dual Graphs&$\sum_{\{m\}}\prod_{v\in V(g)}\prod_{I\in E(g)}(m_I+1)\alpha_{\{m_v\}}$\\
    \raisebox{-0.6cm}{\includegraphics[scale=0.7]{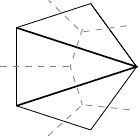}}&\raisebox{-0.6cm}{\includegraphics[scale=.7]{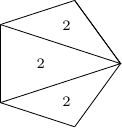}}\quad\raisebox{-0.6cm}{\includegraphics[scale=.7]{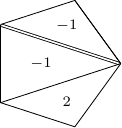}}\quad\raisebox{-0.6cm}{\includegraphics[scale=.7]{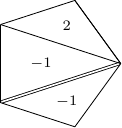}}\quad\raisebox{-0.6cm}{\includegraphics[scale=.7]{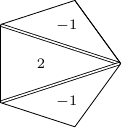}}&$8+4\times2+8=4!$\\
    \raisebox{-0.6cm}{\includegraphics[scale=.7]{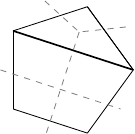}}&\raisebox{-0.6cm}{\includegraphics[scale=.7]{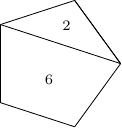}}\quad\raisebox{-0.6cm}{\includegraphics[scale=.7]{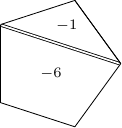}}&$12+12=4!$
\end{tblr}
\caption{All possible dual graphs for each seed and the total of vertex functions at 5-point level.}
\label{5pt seeds}
\end{table}
\begin{longtblr}[
  caption = {All possible dual graphs for each seed and the total of vertex functions at 6-point level.},
  label={6pt_seeds},
]{rowspec={|[1pt]Q[c,m]|Q[c,m]|[dotted]Q[c,m]|[dotted]Q[c,m]|[dotted]Q[c,m]|[dotted]Q[c,m]|[dotted]Q[c,m]|[dotted]Q[c,m]|[1pt]},colspec={Q[c,m]|Q[c,m]|Q[c,m]}}
    Seeds  &Dual Graphs&$\sum_{\{m\}}\prod_{v\in V(g)}\prod_{I\in E(g)}(m_I+1)\alpha_{\{m_v\}}$\\
    \raisebox{-0.6cm}{\includegraphics[scale=.7]{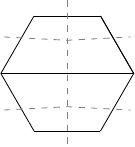}} &\raisebox{-0.4cm}{\includegraphics[scale=.7]{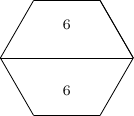}}\quad\raisebox{-0.4cm}{\includegraphics[scale=.7]{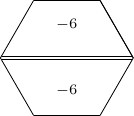}}\quad\raisebox{-0.4cm}{\includegraphics[scale=.7]{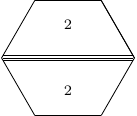}}&$36+72+12=5!$\\
    \raisebox{-0.6cm}{\includegraphics[scale=.7]{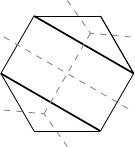}} &\raisebox{-0.4cm}{\includegraphics[scale=.7]{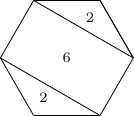}}\quad\raisebox{-0.4cm}{\includegraphics[scale=.7]{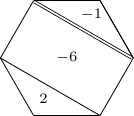}}\quad\raisebox{-0.4cm}{\includegraphics[scale=.7]{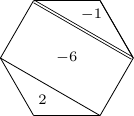}}\quad\raisebox{-0.4cm}{\includegraphics[scale=.7]{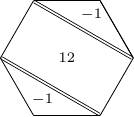}}&$24+2\times24+48=5!$\\
    \raisebox{-0.6cm}{\includegraphics[scale=.7]{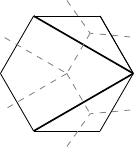}} &\raisebox{-0.4cm}{\includegraphics[scale=.7]{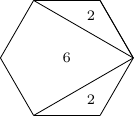}}\quad\raisebox{-0.4cm}{\includegraphics[scale=.7]{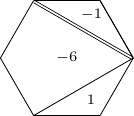}}\quad\raisebox{-0.4cm}{\includegraphics[scale=.7]{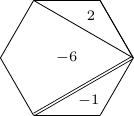}}\quad\raisebox{-0.4cm}{\includegraphics[scale=.7]{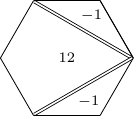}}&$24+2\times24+48=5!$\\\raisebox{-0.6cm}{\includegraphics[scale=.7]{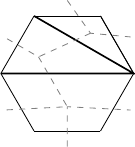}} &\raisebox{-0.5cm}{\includegraphics[scale=.7]{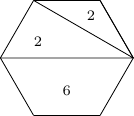}}\quad\raisebox{-0.5cm}{\includegraphics[scale=.7]{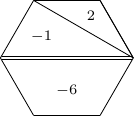}}\quad\raisebox{-0.5cm}{\includegraphics[scale=.7]{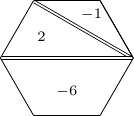}}\quad\raisebox{-0.5cm}{\includegraphics[scale=.7]{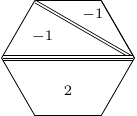}}\quad$\cdots$&$24+24+12+48+12=5!$\\
    \raisebox{-0.6cm}{\includegraphics[scale=.7]{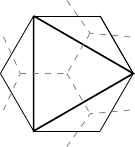}} &\raisebox{-0.5cm}{\includegraphics[scale=.7]{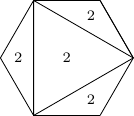}}\quad\raisebox{-0.5cm}{\includegraphics[scale=.7]{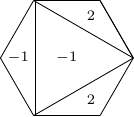}}\quad\raisebox{-0.5cm}{\includegraphics[scale=.7]{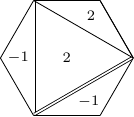}}\quad\raisebox{-0.5cm}{\includegraphics[scale=.7]{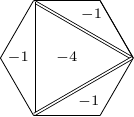}}\quad$\cdots$&$16+3\times8+3\times16+32=5!$\\
    \raisebox{-0.6cm}{\includegraphics[scale=.7]{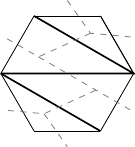}} &\raisebox{-0.5cm}{\includegraphics[scale=.7]{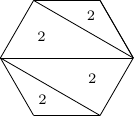}}\quad\raisebox{-0.5cm}{\includegraphics[scale=.7]{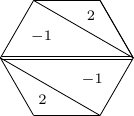}}\quad\raisebox{-0.5cm}{\includegraphics[scale=.7]{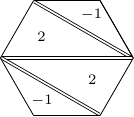}}\quad\raisebox{-0.5cm}{\includegraphics[scale=.7]{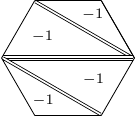}}\quad$\cdots$&$16+3\times8+4+2\times16+32+12=5!$\\
    \raisebox{-0.6cm}{\includegraphics[scale=.7]{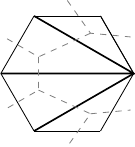}} &\raisebox{-0.5cm}{\includegraphics[scale=.7]{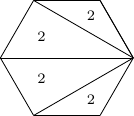}}\quad\raisebox{-0.5cm}{\includegraphics[scale=.7]{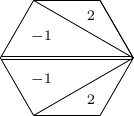}}\quad\raisebox{-0.5cm}{\includegraphics[scale=.7]{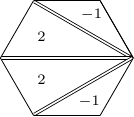}}\quad\raisebox{-0.5cm}{\includegraphics[scale=.7]{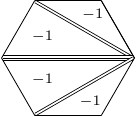}}\quad$\cdots$&$16+3\times8+4+2\times16+32+12=5!$
\end{longtblr}

\end{document}